\documentclass[aps,prb,twocolumn,english, superscriptaddress,showpacs,10pt, reprint]{revtex4-1}

\usepackage{amsmath}
\usepackage{amssymb}
\usepackage{amsfonts}
\usepackage[pdftex]{graphicx}		
\usepackage{array}	
\usepackage{bm}		
\usepackage{color}

\newcommand{\mx}[1]{\ensuremath{\left( \begin{matrix} #1 \end{matrix} \right) }}

\newcommand{\verweis}[1]{\ensuremath{\stackrel{(\ref{#1})}{=}}}

\newcommand{\zref}[1]{(\ref{#1})}
\renewcommand{\vec}[1]{\mathbf{#1}}

\newcommand{\abs}[1]{\left|#1 \right|}
\newcommand{\ring}{\mathring}

\newcommand{\platz}{\ensuremath{\qquad \qquad }}
\newcommand{\intinfty}{\int_{-\infty}^\infty}
\newcommand{\erw}[1]{\langle \, {#1} \, \rangle}
\newcommand{\cl}{\text{cl}}
\newcommand{\q}{\text{q}}
\newcommand{\De}[1]{\stackrel{\Delta}#1 {}\hspace{-1mm}}

\renewcommand{\Im}{\text{Im}\,}
\renewcommand{\Re}{\text{Re}\, }

\newcommand{\sign}{\text{sign}}
\newcommand{\tr}{\text{tr}}

\allowdisplaybreaks
\usepackage{pdfpages}
\usepackage{color}
\usepackage{bm}
\usepackage{xr}
\usepackage[caption=false]{subfig}
\usepackage{graphicx}
\usepackage{hyperref}
\usepackage[figure,table]{hypcap}               
\hypersetup{
pdftitle = {},
pdfsubject = {},
pdfauthor = {},
pdfkeywords = {},
pdfcreator = {},
pdfproducer = {LaTeX with hyperref},
colorlinks = true,
linkcolor = blue, 
anchorcolor = blue,
citecolor = blue, 
filecolor = red, 
menucolor = red, 
pagecolor = red, 
urlcolor  = blue, 
breaklinks = true,
pdfstartview = FitV,
pdfhighlight = /I,
pdfpagelayout = OneColumn,
hypertexnames=true 
}

\renewcommand{\vec}[1]{\boldsymbol{#1}}

\usepackage{color}                                      
\definecolor{d_red}{cmyk}{0.00, 0.81, 1.00, 0.27}
\definecolor{d_orange}{cmyk}{0.00, 0.33, 1.00, 0.00}
\definecolor{d_blue}{cmyk}{0.78, 0.47, 0.00, 0.20}
\definecolor{d_lgreen}{cmyk}{0.07, 0.00, 0.79, 0.29}
\definecolor{d_green}{cmyk}{0.66, 0.00, 0.71, 0.56}
\definecolor{d_blue}{cmyk}{0.78, 0.47, 0.00, 0.20}
\definecolor{d_dblue}{cmyk}{0.91, 0.79, 0.00, 0.22}
\definecolor{d_pink}{cmyk}{0.0, 0.79, 0.37, 0.29}
\definecolor{d_purple}{cmyk}{0.16, 0.54, 0.00, 0.70}
\definecolor{d_paleblue}{cmyk}{0.669, 0.338, 0.00, 0.373}
\definecolor{d_dpaleblue}{cmyk}{0.441, 0.290, 0.00, 0.580}
\definecolor{d_brown}{cmyk}{0.0, 0.490, 0.930, 0.350}
\definecolor{d_turquoise}{cmyk}{0.630, 0.04, 0.0, 0.440}

\newcommand{\av}[1]{\langle #1 \rangle}

\newcommand{\bfk}{{\boldsymbol{k}}}

\newcommand{\bfq}{{\boldsymbol{q}}}

\def\bmx{\begin{pmatrix}}
\def\emx{\end{pmatrix}}

\begin{document}

\title{Luminescence and Squeezing of a Superconducting Light Emitting Diode}

\author{Patrik Hlobil}
\affiliation{Institute for Theory of Condensed Matter, Karlsruhe Institute of Technology (KIT), 76131 Karlsruhe, Germany}
\author{Peter P. Orth}
\affiliation{Institute for Theory of Condensed Matter, Karlsruhe Institute of Technology (KIT), 76131 Karlsruhe, Germany}

\date{\today }

\begin{abstract}
We investigate a semiconductor $p$-$n$ junction in contact with superconducting leads that is operated under forward bias as a light-emitting diode. The presence of superconductivity results in a significant increase of the electroluminescence in a sharp frequency window. We demonstrate that the tunneling of Cooper pairs induces an additional luminescence peak on resonance. There is a transfer of superconducting to photonic coherence that results in the emission of entangled photon pairs and squeezing of the fluctuations in the quadrature amplitudes of the emitted light. We show that the squeezing angle can be electrically manipulated by changing the relative phase of the order parameters in the superconductors. We finally derive the conditions for lasing in the system and show that the laser threshold is reduced due to superconductivity. This reveals how the macroscopic coherence of a superconductor can be used to control the properties of light. 
\end{abstract}
\pacs{74.45.+c, 78.67.-n, 73.40.Lq, 42.50.Dv, 42.55.Px}

\maketitle

\section{Introduction}
Superconductors exhibit quantum coherence of electronic degrees of freedom on a macroscopic scale. For $s$-wave pairing they are characterized by a complex pairing amplitude $\Delta$ with a well-defined phase $\phi$~\cite{Schrieffer-Superconductivity-Book}. This is a consequence of the off-diagonal long-range order present in the two-particle density matrix~\cite{RevModPhys.34.694, Leggett-QuantumLiquids-Book}. The electrons condense into Cooper pairs and form entangled two-electron singlet states. The prospect of harvesting these useful coherence and entanglement properties for the manipulation of quantum states as required for example in quantum information processing and communication~\cite{Bouwmeester:2010:PQI:1965179, OBrien:2009, GisinThew-QuantumCommunication-NatPhot-2007} is a motivation to integrate superconducting elements in semiconductor solid-state nanostructures. Prominent specific goals are the on-demand production of entangled photon pairs due to the recombination of Cooper pairs~\cite{Hanamura-PhysStatSol-2002, Takayanagi-PhysicaC-2010, Suemune-ApplPhysExpress-2010, PhysRevLett.104.156802, PhysRevB.89.094508, Schmidt-CooperPairToPhotons-arXiv-2014, SchroerRecher-SpinEntanglementPnJunction-arXiv-2014} and the generation of non-classical states of light~\cite{PhysRevA.13.2226, zubairy:qo, PhysRevLett.112.077003}. It further brings about the fundamental question of how to efficiently transfer the electronic coherence and entanglement naturally present in a superconductor to excitonic particle-hole pairs in a semiconductor and eventually to photons that are emitted from the hetero-structure.  
\begin{figure}[b!]
\centering
\includegraphics[width=.8\linewidth]{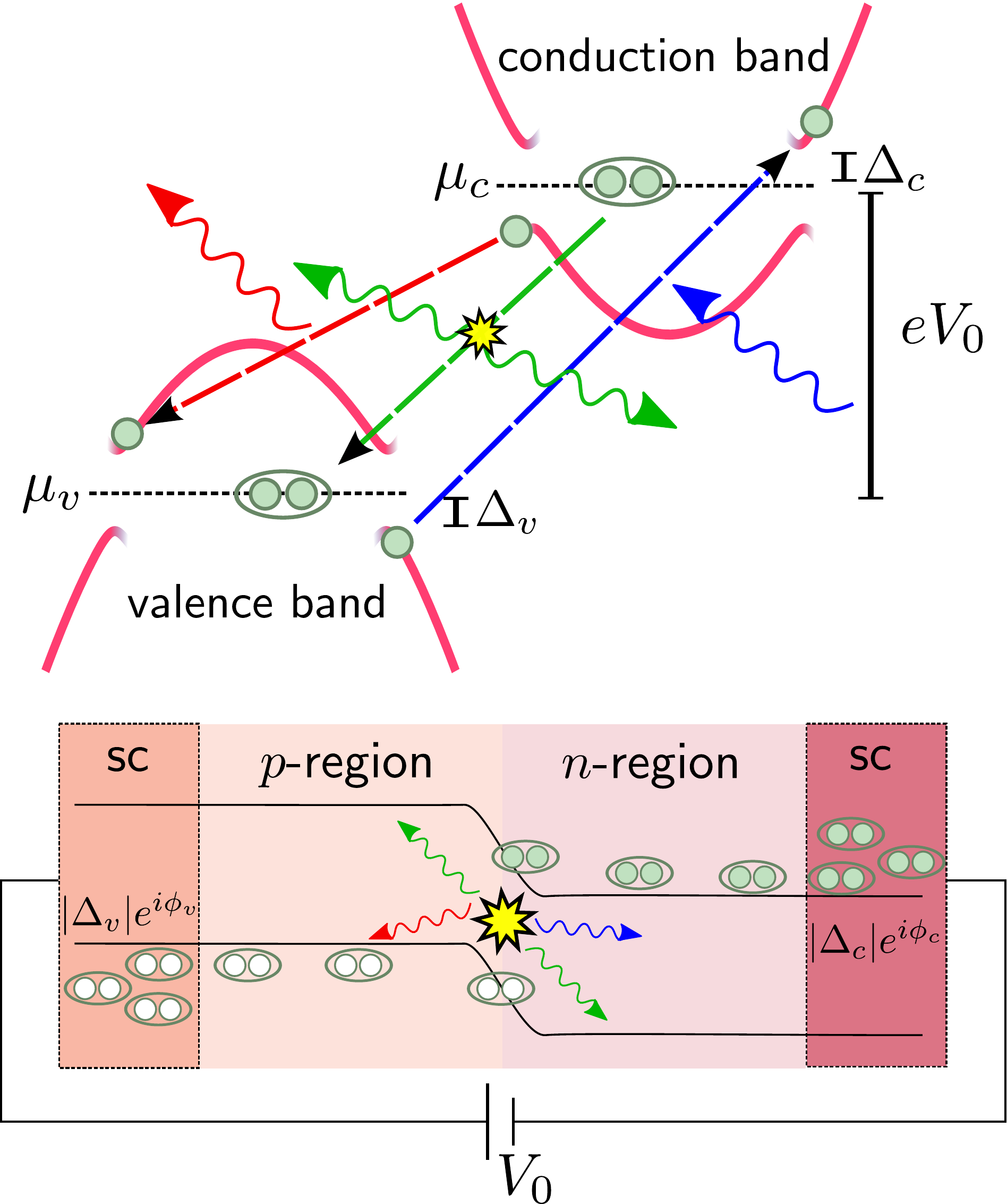}
\caption{(color online). Schematic setup of superconducting light emitting diode with $p$-$n$ junction coupled to superconducting (sc) leads operated under forward bias voltage $V_0$. Basic recombination processes at low temperature $T < |\Delta_c|, |\Delta_v|$: (i) transfer of particle from conduction to valence band upon radiation of a photon (red) with energy $\omega_{\vec q} \leq eV_0 - \abs{\Delta_c}-\abs{\Delta_v}$, (ii) absorption of a photon (blue) with energy $\omega_{\vec q} \geq eV_0 + \abs{\Delta_c} + \abs{\Delta_v}$ upon transfer of a particle from valence to conduction band, and (iii) Cooper pair tunneling upon emission or absorption of two photons (green) with energy $\omega_{\vec q} = e V_0$. The superconducting gaps are denoted $\Delta_v$ ($\Delta_c$) for valence (conduction) band.} 
\label{basicprocesses}
\end{figure}

While the experimental realization of superconductor-semiconductor hybrid nanostructures has proven to be technology challenging~\cite{WeesTakayanagi-ScProximity-Springer}, there have recently been a number of successful experiments using semiconductor nanowires~\cite{Doh08072005, vanDamKouwenhoven-SupercurrentReversalInQuantumDots-Nature-2006}, quantum wells~\cite{Hayashi-ApplPhysExpress-2008,Takayanagi-PhysicaC-2010} and self-organized hetero-superstructures in unconventional superconductors~\cite{Ignatov-arXiv-2013}. Various desired electronic and optoelectronic properties have been observed such as the proximity effect~\cite{Takayanagi-PhysicaC-2010, 0953-2048-23-3-034025} and the Josephson effect~\cite{vanDamKouwenhoven-SupercurrentReversalInQuantumDots-Nature-2006, XiangLieber-MesoscopicJJ-NatureNanotech-2006}, the realization of a superconducting field-effect transistor~\cite{Doh08072005} and enhanced electro- and photo-luminescence rates~\cite{Suemune-JpnJournApplPhys-2006,Idutsu-PhysStatSolC-2009, Suemune-ApplPhysExpress-2010,PhysRevLett.103.187001, PhysRevLett.107.157403,Ignatov-arXiv-2013}. 

Enhanced luminescence rates were observed in a series of experiments using superconductor-semiconductor hetero-structures where a $p$-$n$ junction was contacted with a superconducting Niobium lead on the $n$-side of the junction and cooled below the superconducting transition temperature $T_c$~\cite{Suemune-JpnJournApplPhys-2006,Hayashi-ApplPhysExpress-2008,Idutsu-PhysStatSolC-2009,Suemune-ApplPhysExpress-2010}. In addition, it was clearly shown via the Josephson effect that due to the proximity effect Cooper pairs tunnel into the active region of the $p$-$n$ junction, where they recombine with normal holes~\cite{Takayanagi-PhysicaC-2010}. More recently, enhanced photo-luminescence below a superconducting transition temperature was reported in self-organized superconductor-semiconductor hetero-structures that form naturallly in the iron-based superconductor $\text{K}_x \text{Fe}_{2-y} \text{Se}_2$~\cite{Ignatov-arXiv-2013}. The exciting case where the semiconducting region only contains a few quantum dot states has been studied in Refs.~\onlinecite{PhysRevLett.104.156802,0957-4484-21-27-274004, PhysRevB.87.094511,PhysRevLett.107.073901,PhysRevB.89.094508}.


Here, we investigate a superconductor-semiconductor hetero-structure consisting of a $p$-$n$ junction that is sandwiched between two superconducting leads. The setup is shown in Fig.~\ref{basicprocesses}. We focus on the properties of the light that is emitted from such a superconducting light-emitting diode (LED) under forward bias in the steady-state. To reach a steady-state we consider a coupling of the photons to an external photon bath. One of our main results is a drastic enhancement of the electro-luminescence rate in a sharp spectral window in the presence of superconductivity. This increase, which is shown in Figs.~\ref{Lum1} and~\ref{Lum2} is a consequence of the van-Hove singularities at the edges of the superconducting bands. Similar behavior was observed in a setup with one normal and one superconducting lead in Refs.~\onlinecite{PhysRevLett.103.187001,PhysRevLett.107.157403}. 

We also report the formation of an additional luminescence peak on resonance that is effectively due to transitions of Cooper pairs from the conduction to the valence band. A similar rich peak structure was reported for the case of quantum dots coupled to superconducting leads in Ref.~\onlinecite{PhysRevLett.104.156802}. Further, we analyze the statistical properties of the emitted photons and predict non-classical, two-mode squeezed states of light. Photons with momentum $\vec q$ and $-\vec q$ inherit their squeezing correlations from the electrons forming Cooper pairs. As shown in Fig.~\ref{Squeezing}, the uncertainty of the corresponding two-mode quadrature operators in the squeezed state falls below the minimum uncertainty of coherent states enforced by the Heisenberg limit. Even more importantly, the orientation of the squeezing ellipse can be externally manipulated by changing the relative phase between the two superconductors that are coupled to the $p$-$n$ junction~\cite{PhysRevLett.112.077003}. This shows that one may employ the coherence of a superconductor to tailor the properties of a two-photon quantum state via the transfer of Cooper pair entanglement to the photons. Two-photon correlations in related models are theoretically discussed in Refs.~\onlinecite{PhysRevLett.104.156802, Schmidt-CooperPairToPhotons-arXiv-2014, SchroerRecher-SpinEntanglementPnJunction-arXiv-2014}. Finally, we discuss conditions for lasing in this system and find that the lasing threshold is reduced in the presence of superconductivity, again due to the formation of van-Hove singularities.

The structure of the remainder of the paper is as follows: in the next Sec.~\ref{basmodsec}, we introduce the superconductor-$p$-$n$-superconductor setup and develop the many-body non-equilibrium field theory framework that we use to extract properties of the photons that are emitted in such a device under forward bias. In Sec.~\ref{sec:effect-phot-acti}, we derive the effective photonic action by integrating over the electronic degrees of freedom. We calculate the photonic self-energy due to the coupling to the electrons up to one loop. We discuss in detail the structure and the features of the self-energy both in the presence and in the absence of superconductivity. To obtain a steady state, we consider the leakage of photons out of the system via a coupling to an external photon bath. This yields another contribution to the photonic self-energy. Inverting the Dyson equation, we sum up the complete RPA series and obtain our main analytical result: the dressed photon Keldysh propagators. In the following sections, we extract the physical information they contain about the photonic system. In Sec.~\ref{sec:lumin-sque-sled}, we discuss the luminescence and the light squeezing properties and in Sec.~\ref{sec:steady-state-laser} we derive the conditions to obtain lasing in the system. We determine the lasing threshold also in a physically very transparent way using rate equations. To focus on our main results and the physical consequences in the main text, we shift details of the calculations that yield the effective photon action, the photonic self-energy and the dressed photon propagators to the Appendix.

\section{Hamiltonian and Keldysh action of superconducting LED}   
\label{basmodsec}
\subsection{Hamiltonian of the hetero-structure}
\label{sec:hamiltonian-p-n}
We consider a $p$-$n$ junction that is coupled on each side to a superconducting lead. The system is set under an external forward bias voltage $V_0$ to operate as a LED. Due to the proximity effect Cooper pairs tunnel into the active region of the junction, and participate in recombination processes that lead to the emission of light. We model the active region of the junction together with the photonic degrees of freedom by the Hamiltonian 
\begin{align}
 \label{mod1}
    H &= H_{c}+ H_{v} + H_{el-ph} + H_{ph} + H_{ph}^{bath} \,.
\end{align}
The Hamiltonian of the conduction and the valence electron bands takes the form 
\begin{align}
  \label{mod2}
  H_{\alpha} - \mu_\alpha N_\alpha &= \sum_{\vec k, \sigma=\uparrow,\downarrow} ( \epsilon_{\alpha}( \bfk) - \mu_\alpha ) \alpha_{\vec k\sigma}^\dagger \alpha_{\vec k\sigma} \nonumber \\
&\qquad   + \sum_{\vec k} \bigl(\Delta_\alpha \alpha_{\vec k\uparrow}^\dagger \alpha_{-\vec k \downarrow}^\dagger + \text{h.c.}  \bigr) \,,
\end{align}
where $\alpha^\dag_{\bfk \sigma}$ creates for $\alpha = c (v)$ an electron with momentum $\bfk$ and spin $\sigma$ in the conduction (valence) band. The band dispersions are given by $\epsilon_{\alpha}( \bfk)$ which we will later assume to be of the form $\epsilon_{c}( \bfk) = \bfk^2/2 m_c$ and $\epsilon_{v}( \bfk) = \bfk^2/2 m_v$ with $m_v < 0 < m_c$, \emph{i.e.}, the conduction (valence) band is electron (hole) like. We use $\hbar = 1$ here and in the following. The proximity induced BCS $s$-wave gaps are denoted by $\Delta_\alpha$ and the chemical potentials by $\mu_\alpha$. We have included the chemical potentials into Eq.~\eqref{mod2} to be able to work with time-independent gap functions $\Delta_\alpha$. This requires that we measure the electronic energies with respect to the two different chemical potentials $\mu_\alpha$ in the two respective bands. Their difference is equal to the applied bias voltage 
\begin{align}
  \label{eq:6}
  \mu_c - \mu_v = e V_0\,,
\end{align}
which drives the system out-of-equilibrium. Formally, this is achieved by the gauge transformation $\alpha_{\bfk \sigma} \rightarrow \tilde{\alpha}_{\bfk \sigma} = e^{i \mu_\alpha t} \alpha_{\bfk \sigma}$. This changes the dispersions from $\epsilon_\alpha(\bfk)$ to $\xi_\alpha(\bfk) = \epsilon_\alpha(\bfk) - \mu_\alpha$. 

In addition, this gauge transformation leads to a time-dependence of the electron-photon coupling constant $g_0 \rightarrow g(t) = g_0 e^{i e V_0 t}$ that appears in the coupling Hamiltonian 
\begin{align}
  \label{eq:4}
  H_{el-ph} &= - \sum_{\vec k,\vec k',\sigma}  ( g_0  b_{\vec k-\vec k'} c_{\vec k,\sigma}^\dagger v_{\vec k',\sigma} + \text{h.c.} ) \,.
\end{align}
The operator $b^\dag_{\bfq}$ creates a photon with frequency $\omega_\bfq = c |\bfq|$. The electron-photon coupling is such that the transfer of each electron from the upper conduction to the lower valence band leads to an emission of a single (optical) photon with frequency $\omega_\bfq \approx e V_0$. We restrict ourselves to the case of spin-conserving recombinations. The generalization to circularly polarized photons is straightforward and does not yield any qualitatively new aspects to our model. 

The free photon Hamiltonian is given by 
\begin{align}
  \label{eq:5}
  H_{ph} &= \sum_{\vec q} \omega_{\vec q} b_{\vec q}^\dagger b_{\vec q}\, .
\end{align}
Since photons are constantly produced in the LED due to the presence of the bias voltage, we need to include an external photon bath to deposit the energy and reach a steady-state. This is described by the bath Hamiltonian~\cite{PhysRevLett.96.230602, PhysRevB.75.195331}
\begin{align}
  H_{ph}^{bath} = \sum_{\vec q} \nu_{\vec q} a_{\vec q}^\dagger a_{\vec q} - \sum_{\vec q, \vec q'} \bigl( \lambda_{\vec q \vec q'} b_{\vec q}^\dagger a_{\vec q'} + \text{h.c.}  \bigr) \,.    \label{mod3.1}
\end{align}
The photon bath operator $a^\dag_{\bfq}$ creates a photon with frequency $\nu_\bfq$ in the bath. The coupling constants between system and bath photons are denoted by $\lambda_{\bfq \bfq'}$. We will later integrate over the bath degrees of freedom leading to the possibility to dissipate a (system) photon to the external bath and to create a (system) photon from the bath.

\subsection{Action on the closed time contour}
\label{sec:acti-keldysh-cont}
Due to the applied bias voltage the system that we are considering is in a non-equilibrium state. In the following we want to set up the required non-equilibrium Keldysh field theory formalism that is used to determine observable properties of the system such as the electro-luminescence or the squeezing of the emitted light. The small parameter that controls our (infinite-order) perturbative calculation is the electron-photon coupling constant $g_0$. More precisely, the small dimensionless parameter is given by $|g_0|^2 \rho_F/E_F \ll 1$ for normal conducting leads and $|g_0|^2 \rho_F/|\Delta| \ll 1$ for superconducting leads. Here, $\rho_F$ denotes the electronic density of states at the leads' Fermi energy $E_F$ (see Fig.~\ref{basicprocesses}).  We further assume that we are always below the lasing threshold and derive parametric conditions where lasing occurs later (see Sec.~\ref{sec:steady-state-laser}). 

In non-equilibrium it is required to begin and end the time evolution at the initial state of the system. One thus formulates the field theory on a closed time contour $\mathcal{C}$ such that the action reads~\cite{Schwinger-NonEqSchwingerKeldysh-1961,KonstantinovPerel-NonEqulibriumFieldTheory-JETP, Keldysh-NonEqSchwingerKeldysh-1965,Kamenev-NonEqFieldTheory-Book}
\begin{align}
  \label{eq:1}
  S &= \int_{\mathcal{C}} dt \Bigl[ \sum_{\bfk, \sigma} \Bigl( c_{\bfk,\sigma}^\dag i \partial_t c_{\bfk, \sigma} + v_{\bfk,\sigma}^\dag i \partial_t v_{\bfk, \sigma} \Bigr) \nonumber \\
&\quad + \sum_\bfq \Bigl( b_\bfq^\dag i \partial_t b_\bfq + a_\bfq^\dag i \partial_t a_\bfq \Bigr) - H \Bigr] \,.
\end{align}
Here, $c_{\bfk,\sigma}, v_{\bfk, \sigma}$ denote Grassmann fields and $b_{\bfq}, a_{\bfq}$ denote complex fields arising in a path integral formulation of the action. One can now clearly observe how the applied bias voltage appears by performing the gauge transformation introduced above $\alpha_{\bfk \sigma} \rightarrow \tilde{\alpha}_{\bfk \sigma} = e^{i \mu_\alpha t} \alpha_{\bfk \sigma}$. The time derivative then produces additional terms that contain the chemical potentials such that the energy dispersions are measured with respect to the two different chemical potentials $\epsilon_\alpha(\bfk) \rightarrow \xi_\alpha(\bfk) = \epsilon_\alpha(\bfk) - \mu_\mu$ and $g_0 \rightarrow g(t) = g_0 e^{i e V_0 t}$. We suppress the tilde notation from now on. 

The contour $\mathcal{C}$ starts (and ends) at the initial time $t = 0$, where we assume that $g_0 =0$ and the system is in thermal equilibrium at potentially different temperatures for the electronic and photonic subsystems. At $t>0$ we consider a non-zero electron-photon coupling $g_0$ and a non-zero bias voltage $V_0 > 0$ leading to photon production. In the following, we focus on steady-state properties of the system at times $t, t' \gg t_s$ where $t_s$ is a characteristic time-scale over which transient effects associated with the switch-on decay. We will always check self-consistently that our assumption of a steady-state holds. 

It is convenient to introduce fermionic and bosonic spinors
\begin{align}
\De \Psi_{\vec k}^\zeta &= \mx{ v^\zeta_{\vec k,\uparrow} \\ c^\zeta_{\vec k,\uparrow} \\ (v^\zeta_{-\vec k,\downarrow})^\dagger \\ (c^\zeta_{-\vec k,\downarrow})^\dagger }   \platz \ring \Phi^\zeta_{\vec q} = \mx{ b^\zeta_{\vec q} \\ (b^\zeta_{-\vec q})^\dagger} ,   \label{mod4}
\end{align}
where $\zeta = \pm $ denotes that the time variable is located on the forward $(+)$ or the backward $(-)$ branch of the contour $\mathcal{C}$. The $\Delta$ superscript denotes the fermionic combined Nambu-conduction/valence space and the $\circ$ superscript describes the photon particle-hole space, in which the complex bosons can be described with real spinors, see Eq.~\eqref{eq:39}. It is convenient to make a transformation from the contour $(+,-)$ basis to the RAK (Retarded-Advanced-Keldysh) basis. For bosons this transformation to the classical and quantum fields is given by 
\begin{align}
  \label{eq:9}
 \ring \Phi_\bfq^{cl} &= (\ring{\Phi}^+_\bfq + \ring{\Phi}^-_\bfq)/\sqrt{2} \\ 
\label{eq:10}
 \ring{\Phi}_\bfq^q &= (\ring{\Phi}^+_\bfq - \ring{\Phi}^-_\bfq)/\sqrt{2} \,.
\end{align}
For fermions we follow Larkin and Ovchinnikov~\cite{LarkinOvchinnikov-Keldysh-JETP-1975} and perform the transformation to $(1,2)$ fields as 
\begin{align}
  \label{eq:7}
 \De{\Psi}_\bfk^1 &= (\De{\Psi}_\bfk^+ + \De{\Psi}_\bfk^-)/\sqrt{2} \\
\label{eq:8} 
\De{\Psi}_\bfk^2 &= (\De{\Psi}_\bfk^+ - \De{\Psi}_\bfk^-)/\sqrt{2} \,.
\end{align}
The conjugate Grassmann variable fields $(\De{\Psi}^\zeta_\bfk)^\dag$ are not related to the $\De{\Psi}^\zeta_\bfk$ fields and one may thus choose a different transformation for them 
\begin{align}
  \label{eq:2}
 (\De{\Psi}_\bfk^1)^\dag &= ((\De{\Psi}_\bfk^+)^\dag - (\De{\Psi}_\bfk^-)^\dag)/\sqrt{2} \\
\label{eq:3}
(\De{\Psi}_\bfk^2)^\dag & = ((\De{\Psi}_\bfk^+)^\dag + (\De{\Psi}_\bfk^-)^\dag)/\sqrt{2} \,.
 \end{align}
Combining the two fields as usual we introduce the Keldysh vectors $\hat \Phi_{\vec q} = (\ring \Phi_{\vec q}^\cl ,\ring \Phi_{\vec q}^\q)$ and $\hat \Psi_{\vec k}= (\De{\Psi}_{\vec k}^1 , \De{\Psi}_{\vec k}^2) $. We denote these vectors by $\wedge$ superscripts. We can then write the Keldysh action as 
\begin{align}
S &= \int_{-\infty}^\infty dt dt' \biggl\{ \sum_{\bfk} \hat{\Psi}_\bfk^\dag(t) \hat{G}_{0,\bfk}^{-1}(t,t') \hat{\Psi}_\bfk(t') \nonumber \\ 
& + \frac12 \sum_{\bfq} \hat{\Phi}_\bfq^T(t) \hat{D}_{0,\bfq}^{-1}(t,t') \hat{\Phi}_{-\bfq}(t') \nonumber \\
& + \sum_{\bfk, \bfk'} \hat{\Psi}_\bfk(t) V_{\bfk - \bfk'}(t) \hat{\Psi}_\bfk'(t') \delta(t-t') \biggr\} + S_{ph}^{bath} \,.
\end{align}
The factor of $1/2$ in front of the bosonic propagator arises from our choice of ``real'' bosonic spinors. 
The unperturbed fermionic Green's functions are given by
\begin{align}
\hat G_{0,\vec k}(t,t') &= - i \erw{ \hat \Psi_{\vec k}(t) \hat \Psi_{\vec k}^\dagger(t')  }_0 = {\footnotesize\mx{ \De  G_{0,\vec k}^R(t,t') & \De  G_{0,\vec k}^K(t,t') \\ 0 & \De  G_{0,\vec k}^A(t,t')} } \,.
\end{align}
The retarded and advanced blocks, which only depend on the time difference $\tau = t - t'$, explicitly read
\begin{align}
 &\De  G_{0,\vec k}^{R/A}(t- t') = - i \erw{\De \Psi_{\vec k}^{1/2}(t) [\De \Psi_{\vec k}^{2/1}(t')]^\dagger }_0  \\
 & \quad =  
 \mx{G_{0,\vec k,v}^{(p),R/A} & 0 & P_{0,\vec k,v}^{R/A}  & 0 \\ 
0 & G_{0,\vec k,c}^{(p),R/A}& 0 &  P_{0,\vec k,c}^{R/A}  \\
\bar P_{0,\vec k,v}^{R/A} & 0 & G_{0,\vec k,v}^{(h),R/A} & 0  \\
0 & \bar P_{0,\vec k,c}^{R/A} & 0 & G_{0,\vec k,c}^{(h),R/A}  }_\tau  \nonumber \,.
\end{align}
The average $\av{\; \cdot\; }_0$ is with respect to the free action $S(g_0 = 0)$. 
The particle and hole propagators are denoted by $G^{(p)}$ and $G^{(h)}$. The anomalous electronic propagators which arise due to the presence of superconductivity are denoted by $P$ and $\bar P$. After Fourier transformation $G(\omega) = \int_{-\infty}^\infty d\tau G(\tau) e^{i \omega \tau}$ they take the explicit form
\begin{align}
\mx{G_{0,\alpha}^{(p)}& P_{0,\alpha} \\ \bar P_{0,\alpha} & G^{(h)}_{0,\alpha}}_{\omega,\vec k}^ {R/A} &= \frac{\omega \cdot \openone + \xi_{\alpha}(\vec k) \cdot\sigma_z- \Delta_\alpha \cdot \sigma_+-  \Delta_\alpha^* \cdot\sigma_- }{(\omega \pm i0)^2-\xi_\alpha(\vec k)^2- \abs{\Delta_\alpha}^2}   \label{mod6.1}
\end{align}
with dispersions $\xi_{\alpha}(\bfk) = \epsilon_{\alpha}(\bfk) - \mu_\alpha$ for $\alpha = c,v$ and Pauli matrices $\sigma_{x,y,z}$ where $\sigma_\pm = \frac12 (\sigma_x \pm i \sigma_y)$. 
Since the electrons are assumed to be in thermal equilibrium, one can express the Keldysh Green's function via the fluctuation-dissipation theorem as
\begin{align}
\De  G_{0,\vec k}^{K}(\omega) &= F(\omega) \bigl[ \De  G_{0,\vec k}^R(\omega) -\De  G_{0,\vec k}^A(\omega)   \bigr]. \label{mod6}
\end{align}
with fermionic distribution function $F(\omega) = 1-2 n_F(\omega) = \tanh(\omega/2T_F)$ at temperature $T_F$.

The free Green's function for the photons in the system described by $b_{\bfq}$ can be written as 
\begin{align}
\label{eq:39}
\hat D_{0,\vec q}(t,t') &= - i \erw{\hat \Phi_{\vec q}(t) \hat \Phi_{-\vec q}^T(t') }_0 ={\footnotesize \mx{ \ring D_{0,\vec q}^K(t,t')   &  \ring D_{0,\vec q}^R(t,t')  \\ \ring D_{0,\vec q}^A(t,t')  & 0} }\,.
\end{align}
Note that the expectation value contains the transpose spinor $\Phi_{-\vec q}^T(t')$ (and not the hermitian conjugate). The retarded and advanced blocks take the form
\begin{align}
\ring D_{0,\vec q}^{R/A}(t -t') &= - i  \erw{\ring \Phi_{\vec q}^{\cl/\q}(t) \bigl[\ring \Phi_{-\vec q}^{\q/\cl}(t')\bigr]^T }_0  \nonumber \\
&= \mx{ 0  &  d_{0,\vec q}^{R/A}(t-t')  \\  d_{0,\vec q}^{A/R}(t'-t)  & 0} \,.
\end{align}
where $[d_{0,\vec q}^{R/A}(\omega)]^{-1}= \omega - \omega_{\vec q} \pm i 0$. Since the photonic subsystem is initially in equilibrium, we may write the Keldysh component via the fluctuation-dissipation theorem as
\begin{align}
\ring D_{0,\vec q}^{K}(\omega) &= B_0(\omega) \bigl[\ring D_{0,\vec q}^{R}(\omega)  - \ring D_{0,\vec q}^{A}(\omega) \bigr] \,. \label{mod7}
\end{align}
Here, $B_0$ denotes an initial bosonic distribution function of the uncoupled ($g_0 = 0$) system. As shown below the distribution function $B_0$ will not be important in our calculation of the photon distribution in the steady state. This is rather determined by the interplay between the photon distribution function in the external photon bath and the photon production rate in the LED. 

The electron-photon coupling part in the action contains the vertex expression
\begin{align}
  \label{eq:11}
  V_{\bfk - \bfk'}(t) &= \sum_{\alpha = cl, q} \sum_{i = 1,2} \hat{\gamma}^\alpha \De{g}_i(t) \ring{\Phi}^\alpha_{\bfk - \bfk',i}(t)
\end{align}
with matrices $\hat{\gamma}^{cl} = \hat{\openone}$ and $\hat{\gamma}^q = \hat{\sigma}_x$ in Keldysh space. Using Eq.~\eqref{mod4}, the components of the photon field vector read explicitly $\ring{\Phi}^\alpha_{\bfk - \bfk',1}(t) = b^\alpha_{\bfk - \bfk'}(t)$ and $\ring{\Phi}^\alpha_{\bfk - \bfk',2}(t) = (b^\alpha_{-\bfk + \bfk'}(t))^\dag$. The coupling matrices take the form
\begin{align}
  \label{eq:12}
\De g_1(t)  &= \frac{1}{\sqrt{2}}\mx{ 0 &0& 0 & 0 \\ g(t) & 0 & 0 & 0 \\ 0 & 0 & 0 & -g(t) \\ 0 & 0 & 0 & 0}
\end{align}
and
\begin{align}
 \De g_2(t) &=\frac{1}{\sqrt{2}} \mx{ 0 &\bar g(t) & 0 & 0 \\ 0 & 0 & 0 & 0 \\ 0 & 0 & 0 & 0 \\ 0 & 0 & -\bar g(t) & 0} \,.
\end{align}
with $g(t)= g_0 e^{i eV t}$ and $\bar{g}(t) \equiv (g(t))^*$ denoting the complex conjugate. Finally, the part of the action that describes the coupling of the photons to the photon bath reads
\begin{align}
  \label{eq:13}
  S_{ph}^{bath} &= \int_{\mathcal{C}} dt \biggl[ \sum_{\bfq} a^\dag_{\bfq} (i \partial_t - \nu_\bfq ) a_{\bfq}  \nonumber \\
& + \sum_{\bfq, \bfq'} \bigl( \lambda_{\bfq \bfq'} b^\dag_\bfq a_{\bfq'} + c.c. \bigr) \biggr] \,.
\end{align}
We will later in Sec.~\ref{sec:infl-phot-bath} integrate over the external photon modes $\{a_\bfq\}$ under the standard assumptions of an Ohmic bath in the white noise limit of frequency independent couplings $\lambda_{\bfq \bfq'}$ and bath density of states~\cite{PhysRevB.75.195331}.

\section{Effective photon action}
\label{sec:effect-phot-acti}
In this section, we want to derive an effective action for the photonic degrees of freedom in the system $S_{ph}^{\text{eff}}$ which takes the coupling to the superconducting leads as well as to the photon bath into account. Formally, we integrate over the fermion fields $\hat \Psi_{\vec k}$ and the external bath photon fields $a_{\vec q}$. This integration yields an electronic contribution to the photon self-energy $\ring \Pi_{\vec q}^{\text{el}}(t,t')$ and a bath contribution to the self-energy $\ring \Pi_{\vec q}^{\text{bath}}(t,t')$. 
The effective photon action takes the form 
\begin{align}
S_{ph}^{\text{eff}} &= \frac{1}{2} \sum_{\vec q}\intinfty dt dt' \, \hat \Phi_{\vec q}^T(t)  \biggl[\hat D_{0,\vec q}^{-1}(t,t')  \nonumber \\
& \quad -    \hat \Pi_{\vec q}^{\text{el}}(t,t') -    \hat \Pi_{\vec q}^{\text{bath}}(t,t') \biggr] \hat \Phi_{-\vec q}(t')  \, . \label{seff1}
\end{align}
While we treat the bath self-energy exactly, we take the electronic self-energy up to one loop into account. Importantly, due to the presence of superconductivity there exist non-zero anomalous fermionic Green's functions $P, \bar{P}$. This leads to anomalous terms in the bosonic self-energy $\ring \Pi_{\vec q}^{\text{el}}(t,t')$, which induce similarly anomalous photon expectation values such as $\av{b^\dag_{\bfq}(t) b^\dag_{-\bfq}(t)}$.

In the following, we first derive the photon self-energies and then solve the resulting Dyson equation in the steady state. This corresponds to treating the electron-photon coupling in the random-phase approximation (RPA). It provides us with explicit expressions for the photon propagators - both for normal conducting and for superconducting leads. Details of the calculation are shifted to Appendices~\ref{appepa},~\ref{appelse} and~\ref{appendixdressedpropagators}.
\begin{figure}
\centering
\includegraphics[width=.9\linewidth]{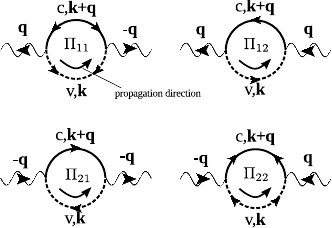}
\caption{Feynman graphs of the one-loop bosonic self-energy $\Pi_{ij,\bfq}^{\text{el}}$ due to the coupling to conduction ($c$) and valence ($v$) electrons, see Eq.~\eqref{eq:16}. External photon propagators (wiggly lines) are not part of the self-energy. Each vertex is associated with a coupling constant $\abs{g_0}$. Wiggly lines denote photons, solid (dashed) lines denote conduction (valence) electron propagators. The anomalous contributions $ \Pi^{\text{el}}_{11},  \Pi^{\text{el}}_{22}$ appear only for superconducting leads $\Delta_c, \Delta_v \neq 0$. }
\label{Seel1}
\end{figure}

\subsection{Electronic feedback}
\label{sec:electronic-feedback}
Let us first focus on the electronic contribution to the effective action in Eq.~\zref{seff1}. It is described by the self-energy $\ring \Pi_{\vec q}^{\text{el}}(t,t')$ and contains the information about the photon absorption and emission processes that involve the transition of electrons between conduction and valence bands.

The electronic part of the photon self-energy has the following structure in the particle-hole basis of $\ring \Phi_{\vec q}$: 
\begin{align}
\ring \Pi_{\vec q}^{\text{el}}(t,t') &= \mx{ e^{i \phi(t+t')} \tilde \Pi_{11,\vec q}^{\text{el}}(\tau)  & \Pi_{12,\vec q}^{\text{el}}(\tau) \\ \Pi_{21,\vec q}^{\text{el}}(\tau) & e^{-i \phi(t+t')} \tilde\Pi_{22,\vec q}^{\text{el}}(\tau)}   \,,
\label{ef1} 
\end{align}
where $\tau = t - t'$ denotes the time difference. 
This structure is identical for the retarded, advanced and the Keldysh components of the self-energy. 
The anomalous components on the diagonal, which are associated with $b_\bfq b_{-\bfq}$ and $b_{\bfq}^\dag b_{-\bfq}^\dag$, depend on the absolute time $t+t'$ via the phase 
\begin{align}
  \label{eq:14}
  \phi (t+t') &= eV_0 (t+t') - \phi_c + \phi_v+ 2 \phi_g \,.
\end{align}
Here, $\phi_{c/v} = \text{arg}(\Delta_{c/v})$ denote the (constant) phases of the superconducting gaps $\Delta_\alpha = |\Delta_\alpha| e^{i \phi_\alpha}$, which depend on microscopic details at the initial time $t=0$, and $\phi_g$ is the phase of the coupling constant $g_0 = |g_0| e^{i \phi_g}$. 
The remaining parts $\tilde \Pi^{\text{el}}_{11,\bfq}$ and $\tilde \Pi^{\text{el}}_{22,\bfq}$, as well as the normal components on the off-diagonals $\Pi^{\text{el}}_{12,\bfq}, \Pi^{\text{el}}_{21,\bfq}$, only depend on the time difference $\tau = t - t'$. They are also independent of the phases $\phi_{c,v,g}$.

We treat these self-energies in the one-loop approximation. As shown in detail in Appendix~\ref{sec:integr-out-electr} the self-energies are given by
\begin{align}
  \label{eq:16}
  &[\Pi^{\text{el}}]_{ij,\vec q}^{\alpha \beta}(t,t') = \\
& \qquad - i \sum_{\vec k} \tr \bigl[  \hat \gamma^\alpha \De g_i(t)     \hat G_{0,\vec k}(t,t')  \hat \gamma^\beta \De g_j(t') \hat G_{0,\vec k+\vec q}(t',t)  \bigr] \nonumber \,.
\end{align}
The corresponding Feynman diagrams are shown in Fig.~\ref{Seel1}. The remaining task is to insert the expression for the fermionic Green's functions, perform the traces over the Keldysh and Nambu indices and to carry out the summation over electronic momenta $\bfk$. 

\subsubsection{Energy scales in the system}
\label{sec:energy-scales-system}
Before we describe the explicit result for the self-energies in the next sections, we want to discuss the different energy scales that are present in the problem. To give a numerical estimate of the different scales, we use realistic values of GaAs. 

There are five important energy scales in the problem: (i) the applied bias voltage $e V_0$, which is of the order of the semiconducting bandgap $D$. This is the largest energy scale in the problem. It sets the photon energy $\hbar \omega_\bfq \approx e V_0$ and the photon momentum $\bfq = e V_0/ \hbar c$, where $c$ denotes the speed of light. For GaAs this energy scale is given by $\hbar \omega_\bfq = 1.424$ eV;
(ii) the semiconductor Fermi energy $\epsilon_F$. It is determined by the effective masses $m_c, m_v$ and the carrier density $n_c,n_v$ as $\epsilon_{F,\alpha} = (3 \pi^2)^{2/3} \hbar^2 n_\alpha^{2/3}/(2 m_\alpha)$. In GaAs one finds $m_c = 0.067 m_e$ and two hole bands with $m_{v,1} = 0.45 m_e$ and $m_{v,2} = 0.082 m_e$, where $m_e$ is the bare electron mass. Typical values for the carrier density are $n = 10^{16} \text{cm}^{-3} - 10^{19} \text{cm}^{-3}$. For this range of doping and approximately using $m_c = - m_v = 0.067 m_e$, we find $\epsilon_F = 2.5 \cdot 10^{-3} - 2.5 \cdot 10^{-1}$ eV. The Fermi velocity follows to $v_{F} = \sqrt{2 \epsilon_F/m} = 4 \cdot 10^{-4} c - 4 \cdot 10^{-3} c$;
(iii) the (proximity induced) superconducting gap $|\Delta|$ for which we estimate $|\Delta| \approx 1 \text{meV}$;
(iv) the strength of the electron-photon coupling $|g_0|^2 \rho$ with conduction band density of states $\rho$. For the range of carrier densities above one finds the estimate $|g_0|^2 \rho = 8.3 \cdot 10^{-6}\text{eV} - 8.3 \cdot 10^{-5}$ eV~\cite{PhysRevLett.112.077003}. This is the smallest energy scale in the system, which justifies our perturbative approach; 
(v) the coupling to an external photon bath induces a photon decay rate $\eta$ (for each photon mode $\bfq$). Our assumption of a steady-state with a finite photon number requires that $\eta$ must be larger than the LED photon production rate. 
 
A result of this estimate is that one cannot neglect the photon momentum $\bfq$. Although it is much smaller than typical electronic momenta $|\bfq|/|\bfk_F| = e V_0 c/(v_F m_c c^2) \approx 0.01 - 0.1$, its associated energy scale is of the same order as the superconducting gap
\begin{align}
  \label{eq:15}
  v_F |\bfq|\approx |\Delta| \,.
\end{align}
In the following, we discuss the explicit result for the electronic part of the photon self-energy $\ring{\Pi}^{\text{el}}_\bfq(t,t')$. We first describe the case of normal conducting leads and then the case of superconducting leads. Taking a finite photon momentum into account serves as a physical cutoff for divergences that would otherwise occur in the imaginary part for the superconducting case. 

\begin{figure}
\centering
\includegraphics[width=\linewidth]{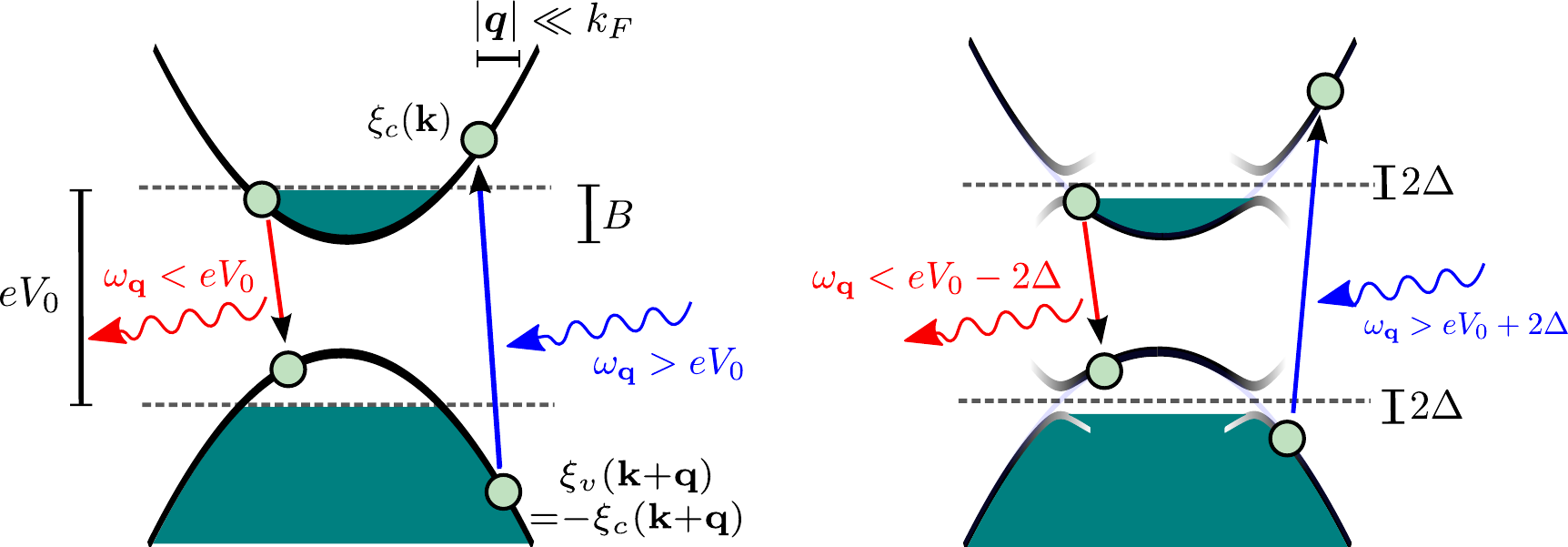}
\caption{(color online). Schematic of the symmetric electronic band dispersion model for normal conducting (left) and superconducting leads (right). Normal state conduction and valence band dispersion obey $\xi_c(\vec k) = - \xi_v(\vec k)$, $v_F$ denotes the Fermi velocity, $k_F$ the Fermi momentum, $B$ the filling factor of the bands and $e V_0$ is the applied bias voltage. Electronic transitions involving photon emission (absorption) are possible at photon energies $\omega_\bfq < e V_0$ ($ \omega_\bfq > e V_0$). Photon momentum $\bfq$ is properly taken into account and results in electronic transitions that are not vertical. In the presence of superconductivity the electrons at the Fermi energies are gapped out, resulting in allowed transitions for $\abs{\omega_\bfq - e V_0} > 2 \Delta$.}
\label{bandsnc}
\end{figure}

\subsubsection{Normal conducting leads}   
\label{senc}
Let us first discuss the electronic contribution to the photon self-energy $\ring{\Pi}^{\text{el}}_\bfq(t,t')$ in the case of normal conducting leads. In this case, the diagonal elements in Eq.~\eqref{ef1}, which are proportional to the product $\Delta_c \Delta_v$, vanish: $\tilde{\Pi}^{\text{el}}_{11,\bfq}(t,t') = \tilde{\Pi}^{\text{el}}_{22,\bfq}(t,t') = 0$.
\begin{figure}[t!]
\includegraphics[width=\linewidth]{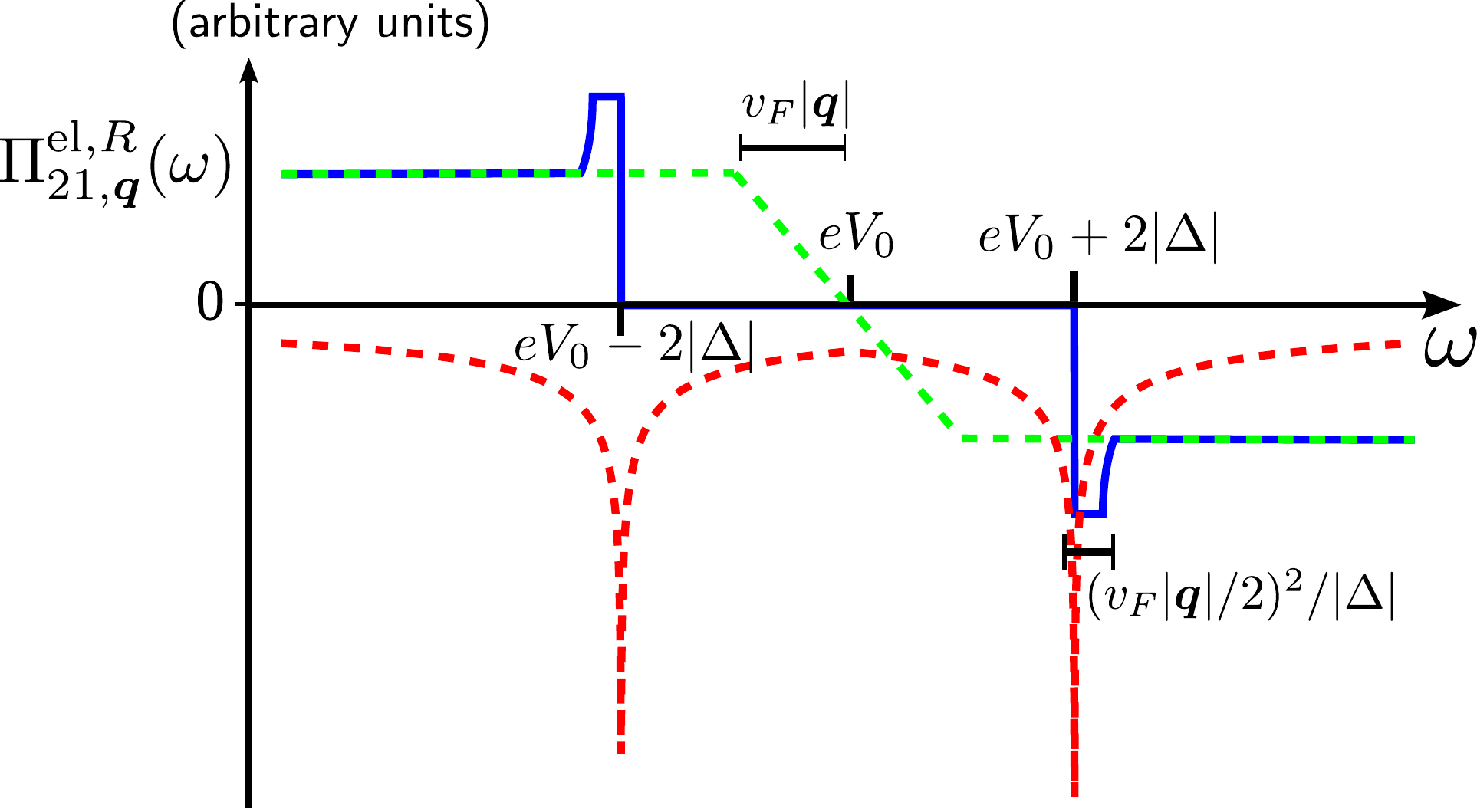}
\caption{(color online). Normal retarded photon self-energy $\Pi_{21,\vec q}^{\text{el},R}(\omega)$ at fixed momentum $\bfq$ as a function of frequency $\omega$. For normal conducting leads the imaginary part is shown in dashed green. The self-energy changes sign at $\omega = e V_0$; positive (negative) sign corresponds to photon production (absorption). For superconducting leads the imaginary part (blue solid) exhibits jumps at $\omega = e V_0 \pm 2 |\Delta|$ (here $|\Delta_c| = |\Delta_v| \equiv |\Delta|$), remains constant over an interval of width $(v_F |\bfq|^2/2)^2/|\Delta|$ and then falls off towards the normal state solution. For superconducting leads the real part (red dashed) exhibits logarithmic divergences at the position of the jumps in the imaginary part. }
\label{NormalSE}
\end{figure}

To be able to perform the summation over momenta in the off-diagonal elements analytically, we make the symmetric choice of assuming parabolic conduction and valence bands bands with effective masses $m_c = - m_v$ and equal superconducting gap amplitudes $|\Delta_c| = |\Delta_v| \equiv |\Delta|$. As a result, the electronic dispersions fulfill $\xi_c(\vec k) = - \xi_v(\vec k)$. The fermionic density of states (DOS) $\rho_c(\omega) = \rho_v(\omega) \equiv \rho(\omega)$ is only weakly energy dependent around the Fermi energy, where $\rho(0) = \rho_F$ corresponds to the DOS at $\mu_{c,v}$. This situation is schematically depicted in Fig.~\ref{bandsnc}. For general dispersions the summation can easily be performed numerically. As shown in detail in Appendix~\ref{appelse}, one finds for the retarded and advanced self-energies at $T_F = 0$
\begin{align}
\Pi_{21,\vec q}^{\text{el},R}(\omega) &= -  i \pi \abs{g_0}^2 \rho \biggl(\frac{\omega_-}{2} \biggr)\begin{cases}  \frac{\omega_-}{v_F |\bfq|} & \text{for}\abs{\omega_-}< v_F |\bfq|  \\ \text{sign}(\omega_-)  & \text{for} \abs{\omega_-}>v_F |\bfq| \end{cases} \nonumber  \\
\Pi_{12,\vec q}^{\text{el},R}(\omega) &= \Pi_{21,\vec q}^{\text{el},A}(-\omega)  \, .   \label{nse1}
\end{align}
Here, the frequencies $\omega_\pm = \omega\pm e V_0$ are measured relative to the applied voltage. 
The retarded function is shown in Fig.~\ref{NormalSE}. Finite fermionic temperatures $T_F > 0$ will smear the zero temperature results, but do not yield qualitatively different results. As the fermions are assumed to be in equilibrium at temperature $T_F$, the corresponding Keldysh self-energies are given by
\begin{align}
\begin{split}
\Pi_{21,\vec q}^{\text{el},K}(\omega) &= \coth\Bigl(\frac{\omega_-}{2T_F} \Bigr) \Bigl[\Pi_{21,\vec q}^{\text{el},R}(\omega)-\Pi_{21,\vec q}^{\text{el},A}(\omega)   \Bigr]  \\
\Pi_{12,\vec q}^{\text{el},K}(\omega) &= \coth\Bigl(\frac{\omega_+}{2T_F} \Bigr) \Bigl[\Pi_{12,\vec q}^{\text{el},R}(\omega)-\Pi_{12,\vec q}^{\text{el},A}(\omega)   \Bigr] \, .    
\end{split}   \label{eq:18}
\end{align}
 As can be seen easily the Keldysh self-energies obey the symmetry $\Pi_{12,\vec q}^{\text{el},K}(\omega) = \Pi_{21,\vec q}^{\text{el},K}(-\omega)$. In Eq.~\eqref{nse1} we have neglected the real part which depends only weakly on frequency. It gives an unimportant renormalization of the photon resonance frequency $\omega_{\vec q}$. The imaginary part of the self-energy $\Pi_{21,\vec q}^R(\omega)$ describes the production and decay of photons due to the coupling to the electrons. From Fig.~\ref{bandsnc} we see that for $T=0$ there are two possible transitions: (i) Photons with energy $\omega_\bfq < eV_0$ are emitted due to transitions of an electron in the conduction band to the valence band, or (ii) photons with energy $\omega_\bfq > eV_0$ are absorbed by raising an electron from the valence band to the conduction band. The absorption is associated with a negative imaginary part of the retarded self-energy in Eq.~\eqref{nse1}, while the emission is associated with a positive imaginary part. The linear dependence of $\Pi_{21,\vec q}^{\text{el},R}(\omega)$ around $\omega = e V_0$ arises due to the restricted phase space of decay and absorption processes for photons with energy $\abs{\omega_\bfq - e V_0} < v_F |\bfq|$. For larger energies $\abs{\omega_\bfq - e V_0} > v_F |\bfq|$ on the other hand, the self-energies reach a constant value because we linearize around the Fermi energies.

\begin{figure}[t!]
\includegraphics[width=\linewidth]{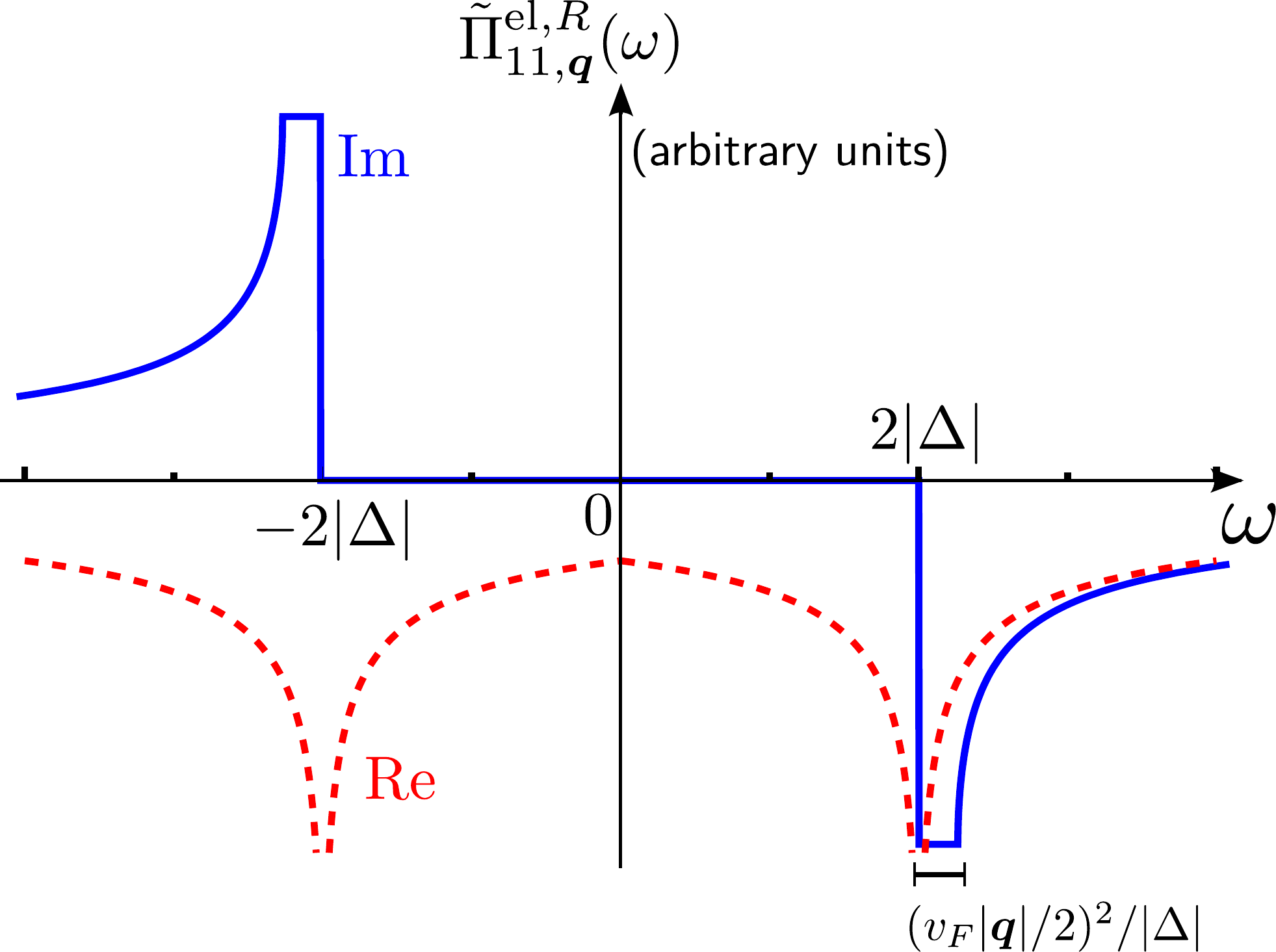}
\caption{(color online). Anomalous retarded photon self-energy $\tilde \Pi_{11,\vec q}^{\text{el},R}(\omega)$ as a function of frequency $\omega$. Real part (red dashed) and imaginary part (blue solid) are non-zero only for superconducting leads and exhibit features similar to the normal self-energy (see Fig.~\ref{NormalSE}). Note that features occur around $\omega = 0$ here. } 
\label{AnomalousSE}
\end{figure}

We have seen that the production of photons is described by a retarded self-energy $\Pi_{21,\vec q}^R(\omega)$ with a positive imaginary part for frequencies $ 0 < \omega < e V_0 $. It changes sign at the applied bias voltage $e V_0$. If there was no further contribution to the total photon self-energy this would be nonphysical, since it does not correspond to the analytic structure required by a retarded bosonic self-energy, which must have a negative imaginary part for $\omega > 0$. Physically, this corresponds to the fact that the assumption of a steady-state breaks down if photons are produced at a constant rate but there is no photon decay mechanism considered. The photon number would simply grow without bound. Therefore, in Sec.~\ref{sec:infl-phot-bath} we take a coupling of the photons to an external photon bath into account. The associated bath induced photon self-energy $\Pi^{\text{bath}}_{ij, \bfq}(\omega)$ will be added to the electronic contribution to the self-energy. The total self-energy will then have a negative imaginary part at positive frequencies $\omega > 0$ as required.

\subsubsection{Superconducting leads}
\label{sec:superconductor}
In the superconducting state the anomalous diagonal elements of the photon self-energy in Eq.~\eqref{ef1} are non-zero. They depend on the total time $T = (t + t')/2$ and it is thus convenient to perform a Wigner transformation $f(\omega,T) = \intinfty d\tau f(\tau,T) e^{i \omega \tau}$, which yields
\begin{align}
\ring \Pi_{\vec q}^{\text{el},R}(\omega,T) &= \mx{ e^{i \phi(2T)} \tilde \Pi_{11,\vec q}^{\text{el},R}(\omega)  & \Pi_{12,\vec q}^{\text{el},R}(\omega) \\ \Pi_{21,\vec q}^{\text{el},R}(\omega) & e^{-i \phi(2T)} \tilde\Pi_{22,\vec q}^{\text{el},R}(\omega)}   \nonumber \\
   \label{nse2} 
\end{align}
with phase $\phi (2 T) = 2eV_0 T + \phi_v - \phi_c + 2 \phi_g$. The self-energy elements obey the relation $\tilde \Pi_{11,\vec q}^{\text{el},R}(\omega)=\tilde \Pi_{22,\vec q}^{\text{el},R}(\omega)$ and $\Pi_{12,\vec q}^{\text{el},R}(\omega)=\Pi_{21,\vec q}^{\text{el},A}(-\omega)$.

As shown in Appendix~\ref{appsup}, the one-loop self-energies at $T_F=0$ take the form
\begin{widetext}
\begin{align}
\label{eq:25}
\tilde \Pi_{11,\vec q}^{\text{el},R/A}(\omega)&=   2\abs{g_0^2} \sum_{\vec k}  \biggl[ \frac{u_{\vec k,v} v_{\vec k,v}u_{\vec k+\vec q,c} v_{\vec k+\vec q,c}}{\omega-E_{v}(\vec k)-E_{c}(\vec k+\vec q)\pm i0} -\frac{u_{\vec k,v} v_{\vec k,v}u_{\vec k+\vec q,c} v_{\vec k+\vec q,c}}{\omega+E_{v}(\vec k)+E_{c}(\vec k+\vec q)\pm i0}           \biggr]  \\
\Pi_{21,\vec q}^{\text{el},R/A}(\omega)&=   2 \abs{g_0^2}\sum_{\vec k}  \biggl[ \frac{v_{\vec k,v}^2 u_{\vec k+\vec q,c}^2 }{\omega_--E_{v}(\vec k)-E_{c}(\vec k+\vec q)\pm i0} -\frac{u_{\vec k,v}^2 v_{\vec k+\vec q,c}^2}{\omega_-+E_{v}(\vec k)+E_{c}(\vec k+\vec q)\pm i0}           \biggr]  \, ,   
\label{nse2.0}
\end{align}
\end{widetext}
where $u_{\vec k,\alpha} = \sqrt{ \frac12  \bigl( 1 + \frac{\xi_{\bfk, \alpha}}{E_{\alpha}(\bfk)}  \bigr) }$ and $v_{\vec k,\alpha} = \sqrt{\frac12  \bigl( 1 - \frac{\xi_{\bfk, \alpha}}{E_{\alpha}(\bfk) } \bigr)}$ are the superconducting coherence factors for the conduction and valence band. They contain the Bogoliubov quasi-particle dispersion relation $E_{\alpha}(\bfk) = \sqrt{ \xi_{\alpha}(\bfk)^2 + |\Delta_{\alpha}|^2}$. 

The normal component is shown in Fig.~\ref{NormalSE} and the anomalous component in Fig.~\ref{AnomalousSE}. 
From the retarded self-energies $\tilde \Pi_{ij,\vec q}^{\text{el},R}(\omega)$ we easily get the Keldysh self-energies as
\begin{align}
\tilde \Pi_{11,\vec q}^{\text{el},K}(\omega)  &= \tilde \Pi_{22,\vec q}^{\text{el},K}(\omega)  = \coth\Bigl(\frac{\omega}{2T_F} \Bigr) \bigl[\tilde \Pi_{11,\vec q}^{\text{el},R}(\omega)-\tilde \Pi_{11,\vec q}^{\text{el},A}(\omega)   \bigr]\,,  \nonumber \\
 \Pi_{21,\vec q}^{\text{el},K}(\omega) &=  \Pi_{12,\vec q}^{\text{el},K}(-\omega)= \coth\Bigl(\frac{\omega_-}{2T_F} \Bigr) \bigl[ \Pi_{21,\vec q}^{\text{el},R}(\omega)- \Pi_{21,\vec q}^{\text{el},A}(\omega)   \bigr]  \,, \label{nse8}   
\end{align}

Let us discuss the retarded self-energies in some detail. The combination $u_{\bfk, \alpha} v_{\bfk, \alpha} = |\Delta_{\alpha}|/(2 E_{\alpha}(\bfk))$ and the diagonal elements $\tilde \Pi_{jj,\vec q}^{\text{el},R/A}$ are therefore proportional to the product $|\Delta_{v}|| \Delta_{c}|$, \emph{i.e.}, they are non-zero only if both leads exhibit superconductivity. In the following, we assume for convenience a momentum independent gap function which is identical for the two bands $|\Delta_{c}| = |\Delta_{v}| \equiv |\Delta|$. Note that the phases of the superconducting order parameters $\phi_c$ and $\phi_v$ are factored out explicitly in Eq.~\eqref{nse2}.

Both diagonal and off-diagonal elements of the self-energy show similar behavior. Their imaginary part vanishes in a region of width $4 |\Delta|$. While the normal components are zero for $|\omega_-| < 2 |\Delta|$ the anomalous components vanish for $|\omega| < 2 |\Delta|$. At the border of these regions the functions exhibit a jump in the imaginary part, then they remain constant over a frequency window of the order of $v_F^2 |\bfq|^2/|\Delta|$ and finally decay towards the result for normal conducting leads (see Eq.~\zref{nse1}) further away from the gapped region. 

Explicitly, the real (imaginary) part of the normal self-energy $\Pi_{21,\vec q}^{\text{el},R}(\omega)$ is (anti)symmetric around the region $\omega=e V_0$. The imaginary part is zero for $|\omega -  e V_0| < 2 |\Delta|$. At the border it exhibits a jump of size 
\begin{align}
\lim_{\delta \rightarrow 0^+} \abs{ \Im \Pi_{21,\vec q}^{\text{el},R}(e V_0+2 |\Delta| + \delta)} = \frac{\pi^2}2 \abs{g_0}^2 \rho_F \frac{|\Delta|}{v_F |\bfq|} \,.   \label{nse5}
\end{align}
Here and in the following we consider $v_F |\bfq| /2 < |\Delta|$, which holds for our realistic choice of parameters (see Sec.~\ref{sec:energy-scales-system}). A jump in the imaginary part yields via the Kramers-Kronig relations a logarithmic divergence in the real part at $\omega= e V_0$. If we define the functions
\begin{align}
 R(w,\chi) &= - \frac{\pi}{2} \abs{g_0}^2  \frac{\rho_F}{\chi}  \ln \biggl[ \sqrt{1+\frac{\chi^2}{\abs{2-\abs{w}}}} + \frac{\chi}{\sqrt{\abs{2-\abs{w}}}}  \biggr] \,,  \nonumber \\
 I(w,\chi) &=  - \frac{\pi}{2} \abs{g_0}^2 \frac{\rho_F}{\chi} \theta(\abs w - 2) \arcsin \biggl[ \frac{\min[\chi,\sqrt{\abs{w}-2} }{\sqrt{\abs{w} - 2}}   \biggr] \,, \label{nse3}
 \end{align} 
the self-energy for frequencies $\omega \approx e V_0 \pm 2 |\Delta|$ can be expressed as
\begin{align}
\Pi_{21,\vec q}^{\text{el},R}(\omega) &= R\Bigl(\frac{\omega_-}{|\Delta|},\frac{v_F \abs{\vec q}}{2 |\Delta|} \Bigr)+ i\,  \text{sign}(\omega_-) I\Bigl(\frac{\omega_-}{|\Delta|},\frac{v_F \abs{\vec q}}{2 |\Delta|} \Bigr) \label{nse4} 
\end{align}
Since $\abs{g_0}^2 \rho_F \ll |\Delta|$ (see Sec.~\ref{sec:energy-scales-system}), the real part is only important around its logarithmic divergence. It quickly decays towards the normal conducting result away from the resonance. The imaginary part of $\Pi^{\text{el}}_{21}$ is given by Eq.~\eqref{nse4} close to the jump and also approaches the constant result that we found for normal conducting leads (see Eq.~\eqref{nse1}).

 The vanishing of the imaginary part for $|\omega_-| < 2 |\Delta|$ can easily be understood if we look at Fig.~\ref{basicprocesses}. For zero temperature there is no possibility for one photon to be either absorbed or emitted within this energy range $\{e V_0 - 2 |\Delta|, e V_0 + 2 |\Delta|\}$ due to the superconducting gaps in the conduction and the valence bands. On the other hand, in a superconductor the DOS diverges at energies $\pm |\Delta|$ relative to the Fermi energy and there are thus many electronic states leading to an enhanced emission and absorption of photons coupling to those states. The imaginary part of the self-energy is therefore enhanced in this region compared to the normal conductor.

We note that it is absolutely essential that we take the finite photon momentum $\bfq$ into account. Otherwise, the imaginary part would exhibit a square-root divergence at $|\omega_-| = 2 |\Delta|$. This divergence is cut-off by finite $\bfq$ at $\abs{\abs{\omega_-}-2|\Delta|}=  \frac{( v_F \abs{\vec q}/2 )^2}{|\Delta|} $ leading to a finite jump instead. Since the imaginary part corresponds to the photon production rate in the superconducting LED, this rate would diverge if one neglects the photon momentum.   

 The anomalous diagonal components of the self-energy $\tilde \Pi^{\text{el}}_{jj}$ are shown in Fig.~\ref{AnomalousSE}. Close to $|\omega| = 2 |\Delta|$ they can also be expressed by the functions defined in Eqs.\zref{nse3} as
\begin{align}
\tilde \Pi_{11,\vec q}^{\text{el},R}(\omega) &= R\Bigl(\frac{\omega}{|\Delta|},\frac{v_F \abs{\vec q}}{2 |\Delta|} \Bigr)+ i\,  \text{sign}(\omega) I\Bigl(\frac{\omega}{|\Delta|},\frac{v_F \abs{\vec q}}{2 |\Delta|} \Bigr) \, . \label{nse4b} 
\end{align}
Away from $|\omega| \approx 2 |\Delta|$ the imaginary part decays like $1/\omega^2$ to zero. This is a faster decay than predicted by Eq.~\zref{nse4b}. 

At finite but small temperatures $T_F \ll |\Delta|$, these results acquire small corrections such as as an exponentially suppressed imaginary part in the gapped regions. At higher temperatures the described features are suppressed, but this also leads to the breakdown of superconductivity and one approaches the results for normal conducting leads.

\subsection{Influence of the photon bath}
\label{sec:infl-phot-bath}
To achieve a steady-state in the system it is required that the photons, which are produced by electrons making a transition from conduction to valence band, may also be absorbed. We thus consider the coupling to an external photon bath as a decay mechanism~\cite{PhysRevLett.96.230602,PhysRevB.75.195331}. As shown in Appendix~\ref{couplingtophotonpath}, the coupling to the bath (see Eq.~\eqref{mod3.1}) gives rise to an additional contribution to the photon self-energy, which corresponds to a finite photon lifetime $\tau_{ph}^{-1} = \eta$. If we neglect the unimportant real part, which has no divergent features, the bath induced photon self-energy reads
\begin{align}
\ring \Pi_{\vec q}^{\text{bath},R}(\omega) &= \mx{  0  & - \Pi_{\vec q}^{\text{bath},R}(-\omega) \\ \Pi_{\vec q}^{\text{bath},R}(\omega) & 0 }   \label{pb1}
\end{align}
with
\begin{align}
\Pi_{\vec q}^{\text{bath},R}(\omega) &=- i \pi \abs{\lambda(\omega)}^2 \rho_{\text{bath}}(\omega) \,.  \label{pb2}
\end{align}
We assumed that the coupling $\lambda_{\vec p,\vec p'}= \lambda(\omega_{\vec p})$ depends only on the photon frequency and introduced the DOS of the external photon bath $\rho_{\text{bath}}(\omega)$. The imaginary part is determined by the spectral function of the external bath, which we assume to be of Ohmic form 
\begin{align}
\pi \abs{\lambda(\omega)}^2 \rho_{\text{bath}}(\omega)  &= \eta \, \theta(\omega) \frac{\omega^2}{\omega^2+\Lambda^2} \label{pb3}
\end{align}
with $\Lambda \ll e V_0$. The spectral function is constant $\eta>0$ for $\omega \gg \Lambda$ and a vanishing function for $\omega \rightarrow 0$. The fact that it decays to zero at small frequency is important since otherwise the Keldysh self-energy
\begin{align}
\ring \Pi_{\vec q}^{\text{bath},K}(\omega,T) &= \mx{  0  & \Pi_{\vec q}^{\text{bath},K}(-\omega) \\ \Pi_{\vec q}^{\text{bath},K}(\omega) & 0 }    \label{pb4} 
\end{align}
with component
\begin{align}
\label{eq:19}
\Pi_{\vec q}^{\text{bath},K}(\omega) &= \coth \Bigl(\frac{\omega}{2T_B} \Bigr) \bigl[ \Pi_{\vec q}^{\text{bath},R}(\omega)- \Pi_{\vec q}^{\text{bath},A}(\omega)  \bigr] 
\end{align}
 would diverge as $1/\omega$ for small $\omega$. Here, we have assumed that the external photon bath is in thermal equilibrium with photon temperature $T_B$. One can easily incorporate a different external photon distribution by replacing $\coth(\omega/2T_B)$ with an arbitrary bath distribution function $B(\omega) = 1+ 2 n^{\text{bath}}(\omega)$, where $n^{\text{bath}}(\omega)$ denotes the number of bath photons in a state of energy $\omega$. 

\subsection{Dressed Photon propagators}  
\label{secdp}
To calculate observable quantities such as the luminescence or the statistical properties of the light emitted from the superconducting LED, we need to find the dressed photon propagator $\hat D_{\vec q}$. The photon self-energies $\hat{\Pi}^{\text{el}}$ and $\hat{\Pi}^{\text{bath}}$ determine the dressed photon propagator $\hat D_{\vec q}$ via the Dyson equation, which reads in Keldysh space~\cite{Kamenev-NonEqFieldTheory-Book}
\begin{align}
\bigl( \hat D_{0,\vec q}^{-1}-\hat \Pi_{\vec q} \bigr) \circ \hat D_{\vec q} &= \hat {\openone} \, ,   \label{dpp1}
\end{align}
where $\hat \Pi_{\vec q}= \hat \Pi_{\vec q}^{\text{el}}+ \hat \Pi_{\vec q}^{\text{bath}}$. Here, $\circ$ denotes a convolution in time. Explicitly, this corresponds to three coupled integral equations for the retarded, advanced and Keldysh components of the dressed propagator
\begin{align}
\label{eq:27}
\bigl( [\ring D_{0,\vec q}^{R/A}]^{-1}-\ring \Pi_{\vec q}^{R/A} \bigr) \circ \ring D_{\vec q}^{R/A} &= \ring {\openone}  \\
\label{dpp3}
\bigl( [\ring D_{0,\vec q}^{R}]^{-1}-\ring \Pi_{\vec q}^{R} \bigr) \circ \ring D_{\vec q}^K &= \ring \Pi_{\vec q}^{K} \circ \hat D_{\vec q}^A \,.  
\end{align}
One can omit the infinitesimal component $[\ring D_{0,\vec q}^{-1}]^K\sim i0$, which acts as a regularization in the non-interacting system, since the coupling to the electrons and the bath induces a finite Keldysh self-energy. Since the anomalous components of the self-energy $\hat{\Pi}^{\text{el}}(\tau,T)$ depend on the absolute time $T$, we perform a Wigner transformation with the ansatz for the full propagator (see Appendix~\ref{appendixdressedpropagators} for details) 
\begin{align}
\ring D_{\vec q}^{R,A,K}(\omega,T) &= \mx{e^{- i\phi(2T)} \tilde D_{11,\vec q}^{R,A,K}(\omega) & D_{12,\vec q}^{R,A,K}(\omega)   \\ D_{21,\vec q}^{R,A,K}(\omega) & e^{ i\phi(2T)} \tilde D_{22,\vec q}^{R,A,K}(\omega)}  \, ,   \label{dpp2}
\end{align}
with the phase factor given in Eq.~\eqref{eq:14}. The dependence on the absolute time is similar but of opposite sign to the one found in the self-energy matrices. 
The explicit forms of the full retarded and advanced photon propagators read
\begin{widetext}
\begin{align}
\label{eq:20}
\tilde D_{11,\vec q}^R(\omega) &=\tilde D_{22,\vec q}^R(\omega) = \frac{- \tilde \Pi_{11,\vec q}^R(\omega)}{\bigl[\omega+e V_0-\omega_{\vec q} -\Pi_{21,\vec q}^R(\omega+eV_0) \bigr]\bigl[\omega-e V_0+\omega_{\vec q} +\Pi_{12,\vec q}^R(\omega-eV_0) \bigr]+ \bigl[ \tilde \Pi_{11,\vec q}^R(\omega)\bigr]^2} \\
\label{eq:21}
D_{12,\vec q}^R(\omega) &=D_{21,\vec q}^A(-\omega) = \frac{\omega-2 e V_0+\omega_{\vec q} + \Pi_{12,\vec q}^R(\omega-2eV_0)}{\bigl[\omega-\omega_{\vec q} - \Pi_{21,\vec q}^R(\omega) \bigr]\bigl[\omega-2e V_0+\omega_{\vec q} + \Pi_{12,\vec q}^R(\omega-2eV_0) \bigr]+ \bigl[ \tilde \Pi_{11,\vec q}^R(\omega-V)\bigr]^2}
\end{align}
The Keldysh components are given by
\begin{align}
\label{eq:22}
\tilde D_{11,\vec q}^K(\omega) = -[\tilde D_{22,\vec q}^K(\omega)]^*  &=\tilde D_{11,\vec q}^R(\omega)\bigl[ \tilde \Pi_{11,\vec q}^K(\omega) \tilde D_{11,\vec q}^A(\omega) + \Pi_{12,\vec q}^K(\omega-V) D_{21,\vec q}^A(\omega-V)  \bigr]  \nonumber \\
& \qquad + D_{12,\vec q}^R(\omega+V) \bigl[ \Pi_{21,\vec q}^K(\omega+V) \tilde D_{11,\vec q}^A(\omega) + \tilde\Pi_{22,\vec q}^K(\omega) D_{21,\vec q}^A(\omega-V)  \bigr] \\
D_{12,\vec q}^K(\omega) = -[D_{21,\vec q}^K(-\omega) ]^*  &= \tilde D_{11,\vec q}^R(\omega-V) \bigl[\tilde \Pi_{11,\vec q}^K(\omega-V) D_{12,\vec q}^A(\omega) + \Pi_{12,\vec q}^K(\omega-2V) \tilde D_{22,\vec q}^A(\omega-V) \bigr] \nonumber\\
& \qquad + D_{12,\vec q}^R(\omega)\bigl[ \Pi_{21,\vec q}^K(\omega) D_{12,\vec q}^A(\omega) + \tilde \Pi_{22,\vec q}^K(\omega-V) \tilde D_{22,\vec q}^A(\omega-V)  \bigr]      \label{prop}
\end{align}
\end{widetext}
By inverting the Dyson equation we have summed up the complete RPA series of bubble diagrams with the electronic contribution to the photon self-energy given in Fig.~\ref{Seel1} and the bath contribution given by Eq.~\eqref{pb2}. In fact, the expressions for the dressed propagators are formally exact if the self-energy was known exactly. This follows from the fact that the structure of the self-energy in Eq.~\eqref{nse2} holds to all orders in perturbation theory.. By inverting the Dyson equation we have only employed this general structure. 

From Eq.~\eqref{eq:21} we find that the photon particle propagator $D_{12,\vec q}^R(\omega)$ shows features at $\omega \approx \omega_{\vec q}$ and $\omega \approx e V_0 \pm 2 |\Delta|$. In the case that the photon is off-resonant with the applied voltage, which is determined by the semiconductor bandgap, $\omega_\bfq \not\approx  e V_0 \pm 2 |\Delta|$ the photon can just propagate through the system and is only weakly interacting with the electrons. In contrast, if the photon is resonant $\omega_\bfq \approx e V_0 \pm 2 |\Delta|$, the enhanced spectral weights around $\omega \approx e V_0 \pm 2 |\Delta|$ correspond to photon-exciton bound states or polaritons. Here, a photon can excite a quasi-particle from the superconducting edges of the valence band to the conduction band, which again recombines under emission of a photon. This process may repeat itself an arbitrary number of times thus forming an electron-photon bound state, a polariton. Since the DOS of the two superconducting bands diverge at $\pm |\Delta|$ (measured from the the Fermi energy), only scattering processes with a photon matching the energy difference $\omega_\bfq \approx e V_0 \pm 2 |\Delta|$ give rise to a large effective coupling between photons and electrons and to the formation of polaritons.  

\section{Luminescence and squeezing properties of the superconducting LED}
\label{sec:lumin-sque-sled}
In this section, we investigate the photon luminescence $\mathcal L(\omega_{\vec q})= \erw{b_\bfq^\dagger b_\bfq}$ of the superconducting LED, \emph{i.e}, the number of photons present in the system in the steady state. We consider both the case of normal conducting leads and the one of superconducting leads. Superconductivity leads to a strong enhancement of the luminescence in a frequency window close to $\omega_\bfq = e V_0 - 2 |\Delta|$. We demonstrate that the superconducting LED emits entangled photon pairs and produces squeezed light of frequency $\omega_{\vec q}= e V_0$. The squeezing occurs in certain two-mode quadrature operators of the light field defined below and implies that the fluctuations in one of the quadrature components falls below the minimal uncertainty of coherent states. 

\subsection{Photon luminescence}
The photon luminescence is defined as the expectation value $\erw{b_\bfq^\dagger b_\bfq}$, which can be expressed via the lesser photon Green's function $D_{12,\vec q}^<(t,t') = - i \erw{b_{\vec q}^+(t) \bar b_{\vec q}^-(t')}$ as
\begin{align}
\mathcal L(\omega_{\vec q})= \erw{b_{\vec q}^\dagger b_{\vec q}} = i \intinfty \frac{d\omega}{2\pi}D_{12,\vec q}^<(\omega)    \label{lm1}
\end{align}
The lesser propagator $D_{12,\vec q}^< = \frac12 (D^K_{12,\vec q} - D^R_{12,\vec q} + D^A_{12,\vec q})$ is a linear combination of the retarded, advanced and Keldysh propagators given in Eqs.~\eqref{eq:20}-\eqref{eq:22}. 

\begin{figure}
\centering
\includegraphics[width=\linewidth]{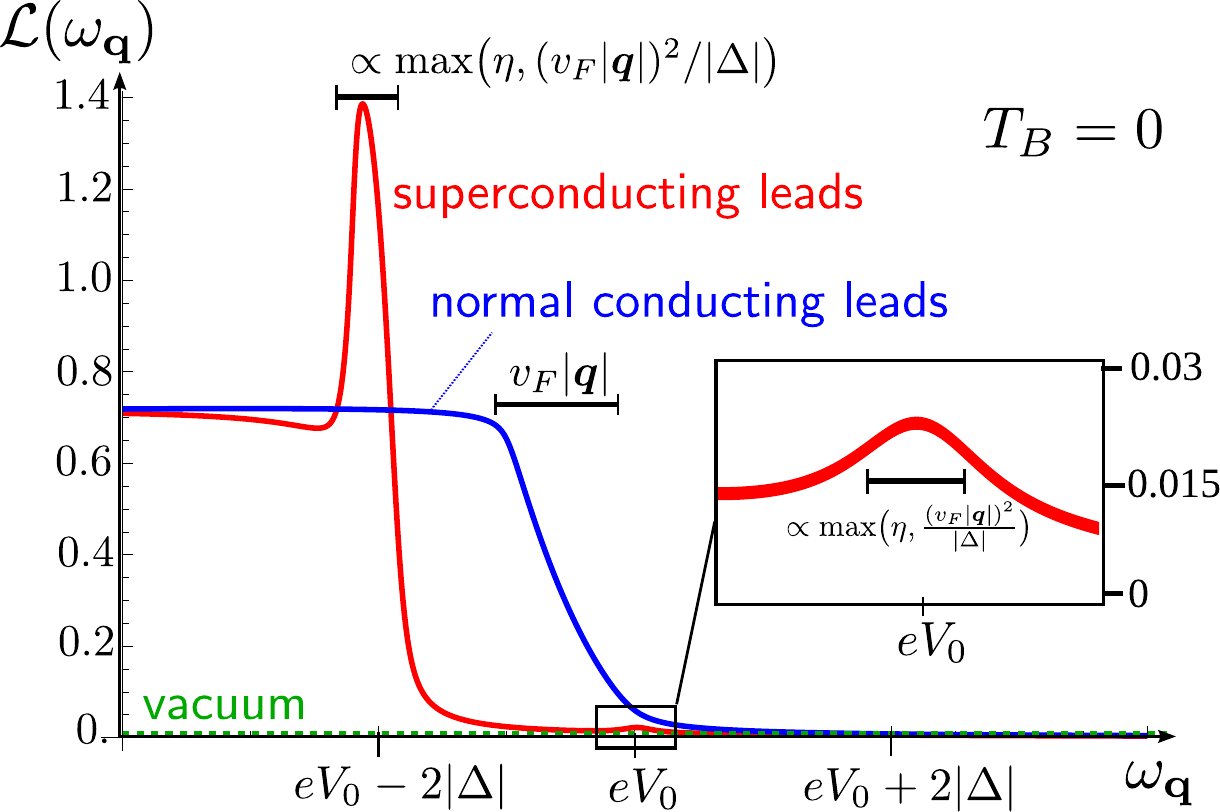}
\caption{(color online). Luminescence $\mathcal L(\omega_{\vec q})= \erw{b_{\vec q}^\dagger b_{\vec q}}$ as a function of photon frequency $\omega_\bfq = c |\bfq|$. We show the luminescence of the LED with normal conducting leads (blue) and superconducting leads (red). Temperature of the external photon bath is kept at $T_B = 0$. Enhanced luminescence around $\omega_\bfq = e V_0 - 2 |\Delta|$ in presence of superconductivity is due to quasi-particles tunneling from conduction to valence band and larger density of states at the edges of the superconducting gap (see Fig.~\ref{basicprocesses}). Inset enlarges region around resonance $\omega_\bfq = e V_0$ showing luminescence peak due to Cooper pair tunneling.}
\label{Lum1}
\end{figure}

\begin{figure}
\includegraphics[width=\linewidth]{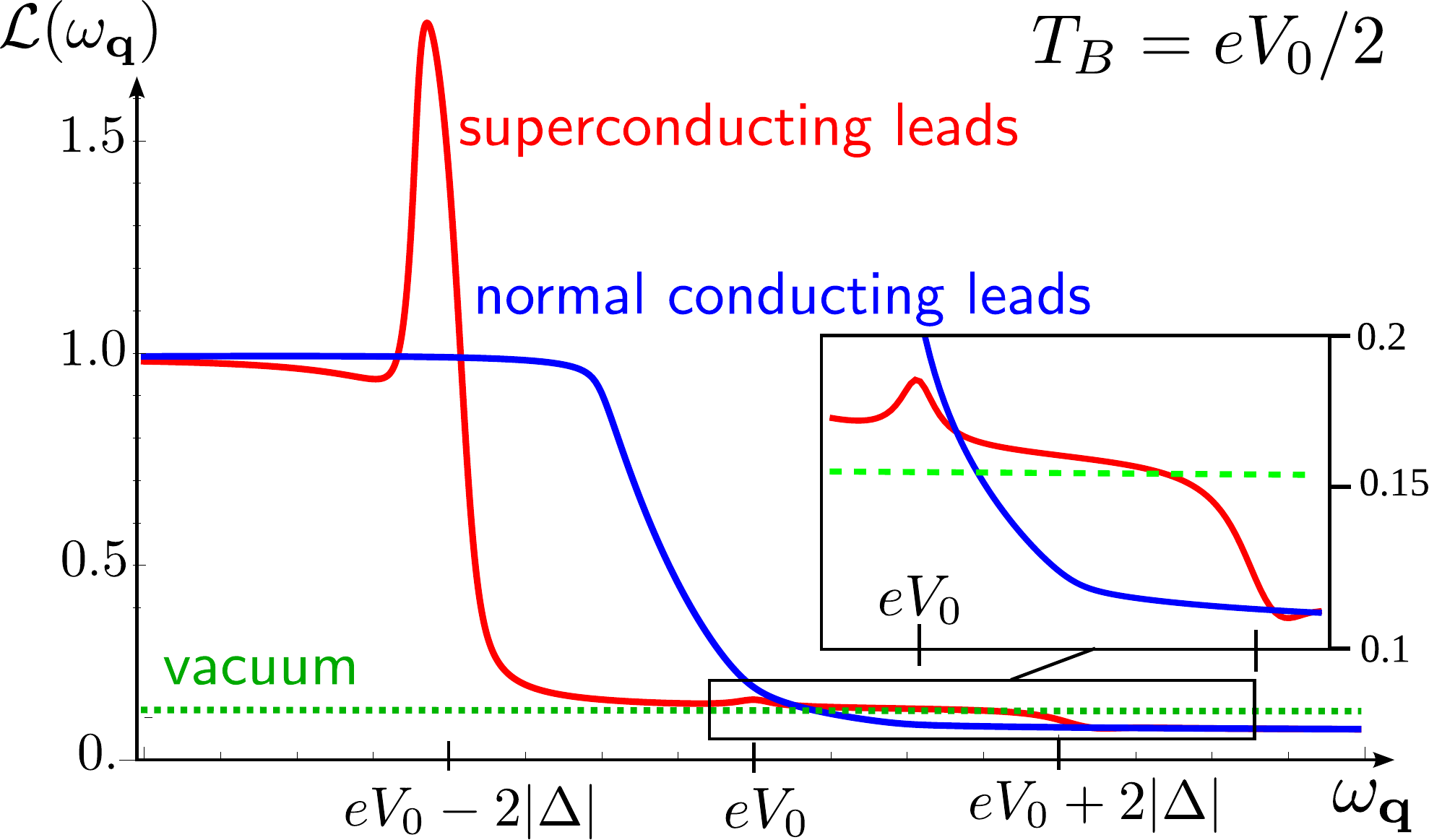}
\caption{(color online). Luminescence $\mathcal L(\omega_{\vec q})= \erw{b_{\vec q}^\dagger b_{\vec q}}$ as a function of photon frequency $\omega_\bfq = c |\bfq|$. External photon bath is kept at temperature $T_B = e V_0/2$. Thermal bath photons (green dashed) with frequency $\omega_\bfq > e V_0 + 2 |\Delta|$ can be absorbed by the $p$-$n$ junction both for normal conducting (blue) and for superconducting leads (red).  Inset enlarges frequency region where the Cooper pair peak and the absorption can be seen. }
\label{Lum2}
\end{figure}

In Fig.~\ref{Lum1} we present the luminescence in the normal and in the superconducting state for an external photon bath at $T_B = 0$. In Fig.~\ref{Lum2} we show the luminescence in the presence of bath photons for a bath kept at temperature $T_B = e V_0/2$. Those photons can be absorbed in the semiconductor junction and transfer electrons from valence to conduction bands. 

Let us first focus the case of $T_B = 0$. For normal conducting leads we then observe that only photons with frequency $\omega_\bfq < e V_0$ are present in the system. This follows from the fact that electronic transitions are only available with energy difference $\omega_\bfq \leq e V_0$ (see Fig.~\ref{bandsnc}). The frequency scale on which the luminescence increases from zero to its constant value is given by the photon momentum $v_F |\bfq|$ with $|\bfq| = e V_0/c$. 
In the case of superconducting leads, we clearly observe a strong enhancement of the number of photons with frequency close to $\omega_\bfq = e V_0 - 2 |\Delta|$. This derives from the large number of electronic states that are pushed out of the gapped region in the superconductor to the border. Photons are produced via a recombination of (Bogoliubov) quasi-particles from the conduction band to the valence band as depicted in Fig.~\ref{basicprocesses}. The behavior of the luminescence can be traced back to the photon self-energy shown in Fig.~\ref{NormalSE}, if one notices that the imaginary parts of the self-energies correspond to photon production rate (for positive imaginary part) and decay rate (for negative imaginary part). 

\begin{figure}[t]
\centering
\includegraphics[width=0.7\linewidth]{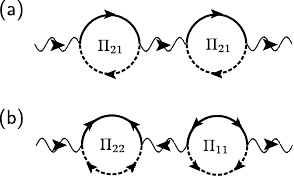}
\caption{Feynman graphs for photon emission processes of order $\mathcal{O}(g_0^4)$ that contribute to the luminescence $\mathcal L(\omega_{\vec q})$. Part (b) contains two anomalous self-energy loops $\Pi_{11}, \Pi_{22}$ and gives rise to the Cooper pair peak seen at $\omega_{\vec q}= e V_0$. }
\label{fig:10}
\end{figure}

In addition to the enhanced photon production at the band edges, the luminescence also exhibits a Cooper pair peak at $\omega_\bfq = eV_0$ in the presence of superconductivity. This stems from Cooper pairs that are transferred from the conduction to the valence band. This process is of the order $|g_0|^4$ and arises from the diagram shown in Fig.~\ref{fig:10} that contains two anomalous self-energy bubbles. This contributions is taken into account in the RPA summation. The luminescence peaks are characterized by a width $\delta \omega = \text{max} \bigl[\eta,  \bigl(\frac{v_F \abs{\vec q}}{2 |\Delta|}\bigr)^2  |\Delta| \bigr]$ which is also the width of the plateau of the imaginary part of the retarded photon self-energies in Eq.~\eqref{nse2.0}.

To obtain the numerical result in Fig.~\ref{Lum1}, we have used parameters that are consistent with our general discussion of energy scales in Sec.~\ref{sec:energy-scales-system}. Specifically, we have expressed all energies in units of the superconducting gap $|\Delta|$ (a realistic value is $|\Delta| = 1 \text{meV}$). We have set $\abs{g_0}^2 \rho_F = |\Delta|/50$, $v_F/c = 0.001$, and $e V_0 = 1000 |\Delta|= 1 eV$. We further assume a simple quadratic conduction band with a band edge that lies at a distance $B=V_0/10$ below the chemical potential $\mu_c$. The electronic density of states thus reads $\rho(\epsilon) = \rho_F \sqrt{1+\epsilon/B}$. For a non-lasing steady-state to exist, the photon decay rate due to the bath $\eta$ must be larger than (see Eq.~\eqref{nse5}) 
\begin{align}
  \label{eq:23}
  \eta_{\text{min}} = \frac{\pi^2 \abs{g_0}^2 \rho_F |\Delta| }{2 (v_F /c) \omega_{\vec q}} \,.
\end{align}
We choose $\eta=1.5 \,  \eta_{\text{min}}$ as even larger decay rates (in particular $\eta > |\Delta|$) result in a photon linewidth that is larger than the superconducting gap which smears the features observed in Fig.~\ref{Lum1}. As soon as $\eta < \eta_{\text{min}}$ the system exhibits lasing and the photon number diverges at a particular frequency. We discuss this possibility in detail in Sec.~\ref{sec:steady-state-laser}. 

If the bath contains real photons they can be absorbed by the hetero-structure. This case is shown in Fig.~\ref{Lum2} where we assume a thermal photon bath at temperature $T_B = e V_0/2$. Photons of frequency $\omega_\bfq \geq e V_0 + 2 |\Delta|$ may be absorbed promoting electrons from the valence to the conduction band. This occurs both for normal conducting as well as for superconducting leads. Here, we neglect the effect on the fermionic distribution and assume that the fermionic distribution functions remains at $T_F = 0$. This can be realized in practice by a fermionic bath of this temperature. In addition to the photon emission peaks, which are unaffected by the presence of bath photons, we now clearly observe an additional absorption dip at $\omega_\bfq = e V_0 + 2 |\Delta|$ in the presence of superconductivity.

\subsection{Light Squeezing}   \label{secsq}
\label{sec:squeezing}
In this section, we investigate the statistical properties of the light that is emitted from the superconducting LED. We find that the photons inherit the anomalous correlations that are present between electrons in a superconductor. In particular, the photon expectation value $\av{b^\dag_{\bfq} b^\dag_{-\bfq}}$ is non-zero, which reduces the quantum fluctuations of certain two-mode quadrature operators of the light field to a value below that found for a coherent state~\cite{zubairy:qo}. The fluctuations of the conjugate operator are of course increased such that the Heisenberg uncertainty limit is obeyed. The superconducting LED therefore emits two-mode squeezed light, which can act as a resource for quantum information processing and metrology~\cite{OBrien:2009, Kimble:1987,LaPorta:1987,Schnabel:2010}.

The light squeezing in this setup was studied in Ref.~\onlinecite{PhysRevLett.112.077003} within lowest order perturbation theory in the electron-photon coupling $g_0$. To this order, the luminescence $\av{b^\dag_\bfq b_\bfq}$ vanishes on resonance $\omega_\bfq = e V_0$ (at $T_B =0$) where the lowest order terms are $\mathcal{O}(g_0^4)$. The amount of squeezing is then solely determined by the anomalous expectation value $\av{b^\dag_{\bfq} b^\dag_{-\bfq}}$. Here, we consider the effect of the diagonal (photon number) expectation values $\av{b^\dag_\bfq b_\bfq}$ on the squeezing by summing the complete RPA series. We make a quantitative prediction for the maximal reduction of quantum fluctuations that is achievable in this setup. 

\begin{figure}
\centering
\includegraphics[width=\linewidth]{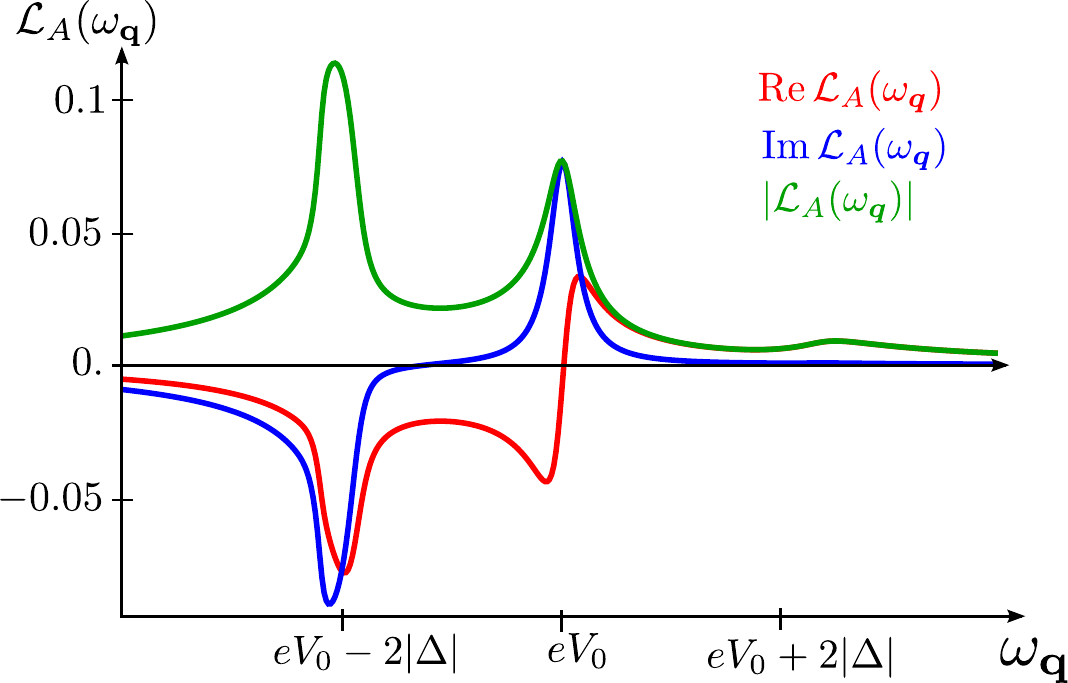}
\caption{(color online). Anomalous photon expectation value $\mathcal L_A(\omega_{\vec q})= e^{i \phi(2t)}\erw{b_{\vec q}(t) b_{-\vec q}(t)}$ as a function of photon frequency $\omega_\bfq = c |\bfq|$. We show the real part (red), imaginary part (blue) and absolute value (green). External bath temperature is set to $T_B = 0$.}
\label{LA}
\end{figure}

Squeezing occurs in the two-mode quadrature operators
\begin{align}
A_{\vec q}^\pm = \mathcal N_\pm \bigl[ \tilde b_{\vec q}^\dagger + \tilde b_{-\vec q}^\dagger \pm \text{h.c.}   \bigr] \, ,   \label{sq1}
\end{align}
where $\mathcal N_+ = 2^{-3/2}$ and $\mathcal N_- = - i 2^{-3/2}$. The photon operators are defined in the rotating frame $\tilde b_{\vec q } = b_{\vec q}  e^{i \omega_{\vec q }t}$. The fluctuations of these two-mode quadrature operators are given by~\cite{PhysRevLett.112.077003}
\begin{align}
\erw{\bigl( \Delta A_{\vec q}^\pm  \bigr)^2} &= \frac{1}{4} \biggl[ 1 + 2 \erw{\tilde b_{\vec q}^\dagger \tilde b_{\vec q}} \pm 2 \Re \erw{\tilde b_{\vec q} \tilde b_{-\vec q}} \biggr] \,,   \label{sq2}
\end{align}
where $(\Delta A)^2 = ( A - \av{A})^2$. The state is truly squeezed if the fluctuations in one of the quadrature operators fall below the Heisenberg uncertainty limit: $\erw{\bigl( \Delta A_{\vec q}^i  \bigr)^2}<\frac{1}{4}$ for either $i=+$ or $i=-$. If the number of photons $\erw{ b_{\vec q}^\dagger  b_{\vec q}}$ in a mode $\bfq$ is zero, a finite anomalous expectation value of $\Re \erw{\tilde b_{\vec q} \tilde b_{-\vec q}}$ always results in squeezing. On the other hand, in the presence of photons one must compare the expectation values $\erw{\tilde{ b}_{\vec q}^\dagger  \tilde{b}_{\vec q}}$ and $\erw{ \tilde{b}_{\vec q}^\dagger  \tilde{b}^\dag_{-\vec q}}$ to each other. Since the number of photons is smaller at low temperature, the resulting squeezing amplitudes are larger. We thus focus on the case of $T_B =0$ in the following. A crucial observation is that close to resonance $\omega_\bfq = e V_0$ it holds that the luminescence $\erw{\tilde{ b}_{\vec q}^\dagger  \tilde{b}_{\vec q}} = \mathcal{O}(g_0^4)$ while the anomalous expectation values are already of order $\erw{ \tilde{b}_{\vec q}^\dagger  \tilde{b}^\dag_{-\vec q}} = \mathcal{O}(g_0^2)$. We obtain the anomalous photon expectation value 
\begin{align}
\label{eq:32}
\erw{\tilde b_{\vec q}(t) \tilde b_{-\vec q}(t)} &= e^{2i \omega_{\vec q} t} \int \frac{d\omega}{2\pi} i D_{11,\vec q}^<(\omega, T=t)  \nonumber \\
&= e^{ i [ 2 \omega_{\vec q}t -\phi(2t)]}  \int \frac{d\omega}{2\pi}i \tilde D_{11,\vec q}^<(\omega)
\end{align}
by an integral over the lesser anomalous propagator $D^<_{11,\bfq} = \frac12 (D^K_{11,\bfq} - D^R_{11,\bfq} + D^A_{11,\bfq})$. We define the anomalous luminescence by $\mathcal{L}_A(\omega_\bfq) = e^{i \phi(2t)} \erw{b_{\vec q}(t) b_{-\vec q}(t)} = \int \frac{d\omega}{2\pi} i \tilde{D}_{11,\vec q}^<(\omega) = e^{i l_\bfq} |\mathcal{L}_A(\omega_\bfq)|$ to arrive at 
\begin{align}
\label{sq4}
\erw{\bigl( \Delta A_{\vec q}^\pm  \bigr)^2} &= \frac{1}{4}\Bigl( 1 + 2 \mathcal L(\omega_{\vec q}) \\
& \qquad \pm 2 \cos\bigl[2(\omega_{\vec q}- e V_0)t + \phi_{\vec q} \bigr]  |\mathcal{L}_A(\omega_{\vec q})| \Bigr) \nonumber \,.  
\end{align}
Here, the initial phase of the last terms reads $\phi_{\vec q} = \arg[ \Delta_v \Delta_c^* g_0^2 \mathcal{L}_A(\omega_{\vec q})  ] = \phi_v-\phi_c + 2 \phi_g + l_{\vec q}$ and the time dependence vanishes for photons on resonance $\omega_\bfq = e V_0$. The anomalous luminescence $\mathcal{L}_A(\omega_\bfq)$ is shown in Fig.~\ref{LA}. At $T_B =0$ it exhibits two main peaks: one at $\omega_\bfq = e V_0 - 2 |\Delta|$ corresponding to the transition of a Bogoliubov quasi-particle from the conduction to the valence band and one at $\omega_\bfq = e V_0$ corresponding to the transition of a Cooper pair. It is important to note that the peak at $\omega_\bfq = e V_0 - 2 |\Delta|$ implies that breaking up (two) different Cooper pairs still leads to the emission of correlated and phase coherent photons due to the presence of macroscopic electronic BCS condensates. Both processes are depicted in Fig.~\ref{basicprocesses}. Due to the finite number of photons at $\omega_\bfq = e V_0 + 2 |\Delta|$ which are produced in the system (see luminescence in Fig.~\ref{Lum1}) we observe a small peak of $\mathcal{L}_A(\omega_\bfq)$ at $\omega_\bfq = e V_0 + 2 |\Delta|$. This peak corresponds to an absorption process of a Bogoliubov quasi-particle from the valence to the conduction band. 

In order to obtain squeezing for the mode with photon momentum $\vec q$, it is required that $\abs{\mathcal L_A(\omega_{\vec q})} > \abs{\mathcal L(\omega_{\vec q})} $. Comparing the normal luminescence $\mathcal{L}(\omega_\bfq)$ in Fig.~\ref{Lum1} with the anomalous luminescence $\mathcal{L}_A(\omega_\bfq)$ in Fig.~\ref{LA}, we find that squeezing is maximal for photons on resonance $\omega_\bfq = e V_0$, which corresponds to the transitions involving Cooper pairs. 

\begin{figure}
\centering
\includegraphics[width=\linewidth]{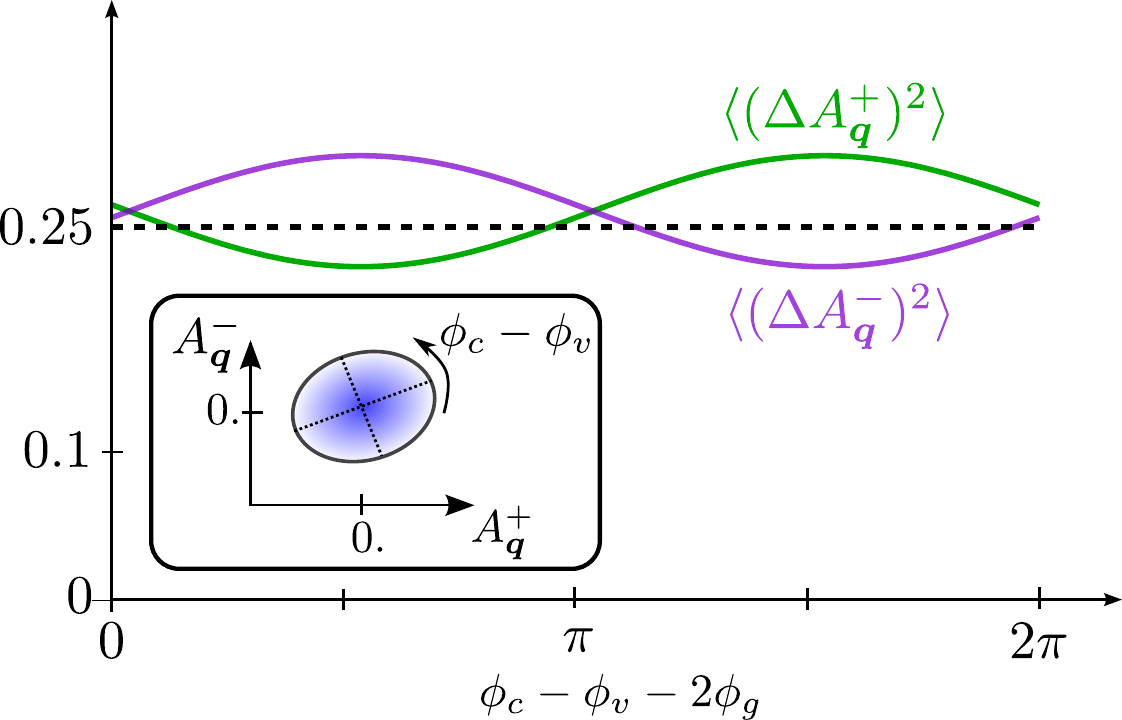}
\caption{(color online). Fluctuations $\erw{\bigl( \Delta A_{\vec q}^\pm  \bigr)^2}$ of the two-mode quadrature operators $A_\bfq^\pm$ for $|\bfq| = e V_0/c$ (on resonance) as a function of superconducting phase difference $\phi_c - \phi_v + 2 \phi_g$. The phase $\phi_g$ denotes the constant phase of the electron-photon coupling constant $g_0$. For a range of phase differences one of the fluctuation amplitudes falls below the Heisenberg uncertainty minimum for the symmetric case $\erw{\bigl( \Delta A_{\vec q}^\pm  \bigr)^2} = 1/4$. This shows that a superconducting LED emits squeezed light. Inset shows mean and uncertainty of two-mode quadrature operators $A_{\vec q}^\pm$. The orientation of the squeezing ellipse is controlled by $\phi_c - \phi_v$. } 
\label{Squeezing}
\end{figure}

In Fig.~\ref{Squeezing} we show the amount of squeezing on resonance at $T_B = 0$. It shows that squeezing can be controlled by the relative phase between the two superconductors $\phi_c - \phi_v$. The fluctuations of either $\Delta A^{+}_\bfq$ or $\Delta A^-_\bfq$ fall below the Heisenberg uncertainty limit for a broad range of relative phases. The maximal amount of squeezing for our realistic choice of parameters is about $10$ percent.

In the experimental setup of the $p$-$n$ junction the relative phase $\Delta \phi = \phi_c - \phi_v$ depends on microscopic details such as the initial switch-on time. In an experiment it will be random from experiment to experiment but fixed within one run. If one places two superconducting $p$-$n$ junctions in parallel in a SQUID geometry one can control the relative phase $\phi_c - \phi_v$ between the two junctions. Since this will essentially change the individual relative phases $\Delta \phi^1$, $\Delta \phi^2$ as well, this provides a way to change $ \phi_c - \phi_v$ with a magnetic field. 

\section{Steady state and lasing threshold}
\label{sec:steady-state-laser}
In our calculation so far we have assumed that the system reaches a steady-state with a finite number of photons in the system. This requires that the bath absorbs the photons that are produced in the LED sufficiently fast. In this section, we derive the exact requirements that have to be fulfilled for this to be the case. Otherwise, the luminescence diverges at certain frequencies and the system exhibits lasing. We discuss the lasing conditions and give the frequency window where lasing occurs. Finally, we calculate the steady-state photon distribution using rate equations.  

\subsection{Lasing condition} 
\label{sec:lasing-condition}
To derive the conditions that are required for a steady-state with a finite photon number to exist, we follow a discussion given in Ref.~\onlinecite{raey}. The key idea is to determine when and for which frequencies the luminescence exhibits a divergence. This corresponds to a transition into a lasing regime. We first consider the case of normal conducting leads and then the one of superconducting leads. 

The luminescence $\mathcal{L}(\omega_\bfq) = \av{b^\dag_\bfq b_\bfq}$ can be obtained via the lesser Green's function $D_{12,\vec q}^< = \frac12 (D^K_{12,\vec q} - D^R_{12,\vec q} + D^A_{12,\vec q})$ (see Eq.~\eqref{lm1}). We can parametrize the inverse retarded, advanced and Keldysh Green's functions as 
\begin{align}
\label{eq:24}
[D_{12,\vec q}^{R/A}(\omega)]^{-1} &= A_{\vec q}(\omega) \pm i B_{\vec q}(\omega)     \\
\label{lm2}
[D_{12,\vec q}^K(\omega)]^{-1} &= i C_{\vec q}(\omega) \,.
\end{align}
Using the matrix structure in Keldysh space (see Eq.~\eqref{eq:39}) one obtains immediately
\begin{align}
\label{eq:26}
 D_{12,\vec q}^{R/A}(\omega) &= \bigl[ A_{\vec q}(\omega) \pm i B_{\vec q}(\omega)\bigr]^{-1} \\
\label{lm2b}
 D_{12,\vec q}^K(\omega) &= - \frac{ [D_{12,\vec q}^K(\omega)]^{-1}}{[D_{12,\vec q}^{R}(\omega)]^{-1}[D_{12,\vec q}^{A}(\omega)]^{-1}} \, .
 \end{align}  
The zeros of $A_{\vec q}(\omega)$ describe the excitations of the system and $B_{\vec q}(\omega)$ their linewidth. Let us assume that there exists a resonance at the renormalized photon frequency $\omega=\omega_{\vec q}^*$ such that $A_{\vec q}(\omega_{\vec q}^*)=0$. It is then required that the imaginary part fulfills $B_{\vec q}(\omega_{\vec q}^*)>0$ to obtain a proper retarded Green's function with the poles lying in  the lower complex frequency plane. 
\begin{figure}
\centering
\includegraphics[width=\linewidth]{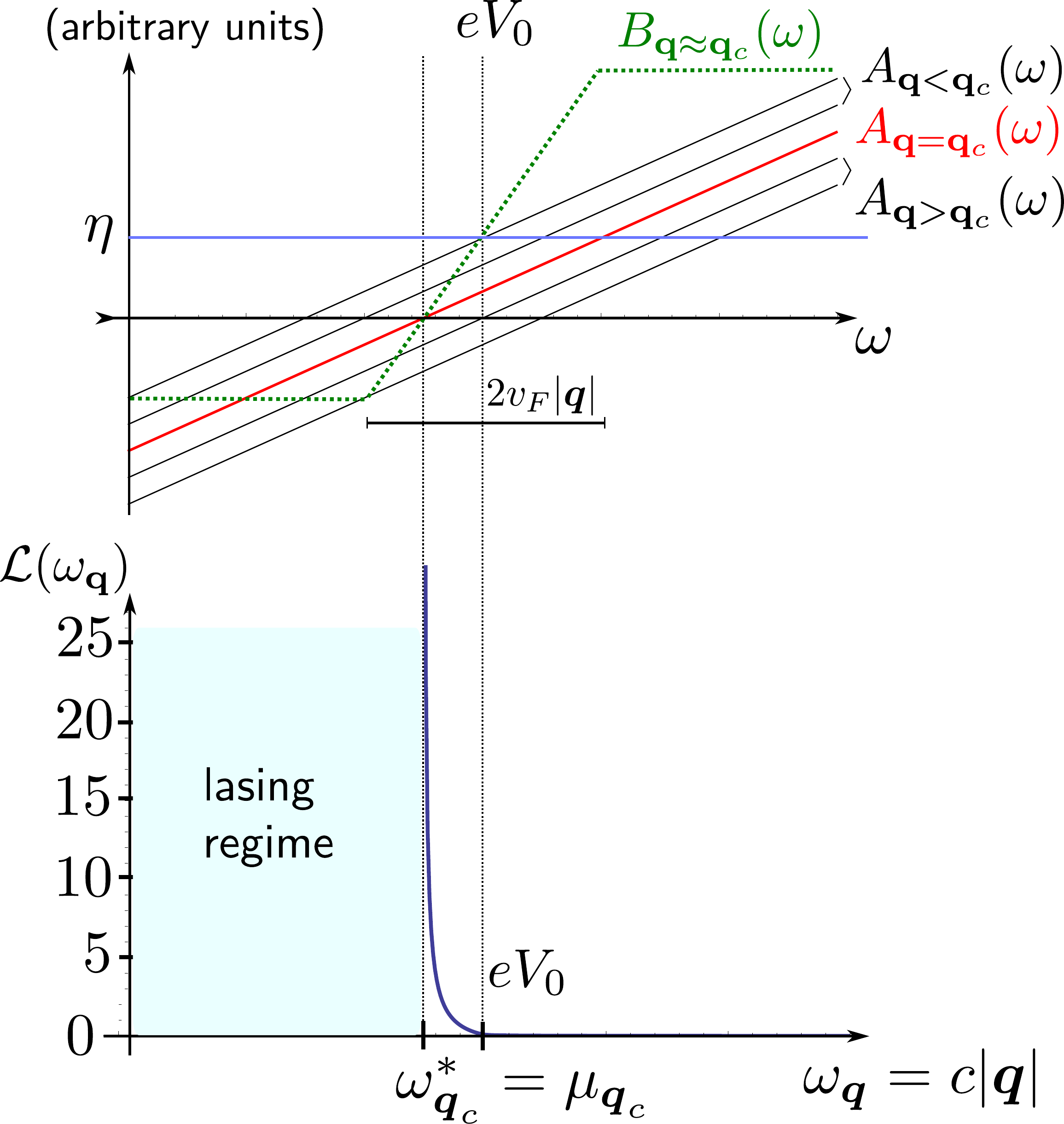}
\caption{(color online). Lasing condition for normal conducting leads. Upper panel shows nearly momentum independent imaginary part $B_{\vec q}(\omega)\approx B_{\vec q_c}(\omega)$ (green dashed) of inverse retarded propagator $[D_{12,\vec q}^R(\omega)]^{-1}$. Lasing occurs for momenta larger than $|\bfq_c| \approx e V_0/c$. Corresponding real part $A_{\vec q}(\omega)$ is shown for different momenta $\vec q$ around $\bfq_c$. The photon bath decay rate is chosen $\eta < \pi \abs{g_0}^2 \rho_F$ to fulfill the laser threshold relation for frequencies $\omega_\bfq \lesssim e V_0$. Lasing occurs for mode with momentum $\bfq = \vec q_c$ where both real part $A_{\vec q_c}(\omega)$ (red) and imaginary part $B_{\vec q_c}(\omega)$ have simultaneous zeros. Lower panel shows the luminescence $\mathcal L(\omega_{\vec q}) = \erw{b_{\vec q}^\dagger b_{\vec q}}$ as a function of photon energy $\omega_{\vec q} = c |\bfq|$. At $\omega_{\vec q} < \omega_{\bfq_c}$ the luminescence diverges, which denotes a violation of our assumption of a (non-lasing) steady-state with finite photon number. In the lasing regime  $\omega_{\vec q} < \omega_{\vec q_c}$ (light blue region) the system produces photons at a faster rate than the absorption $\eta$ due to the bath occurs. }
\label{ans0}
\end{figure}
 \begin{figure}
\centering
\includegraphics[width=\linewidth]{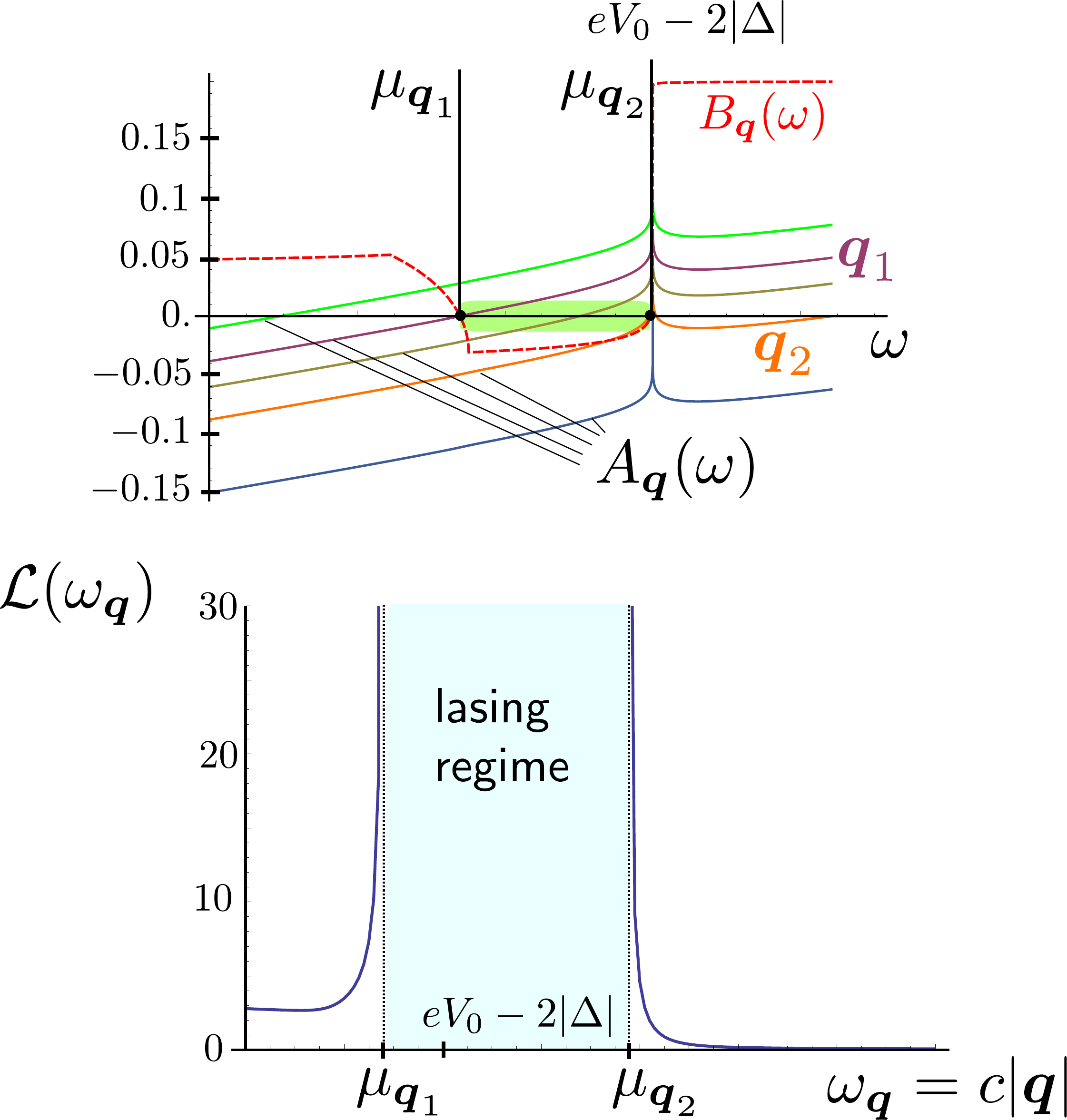}
\caption{(color online). Lasing condition for superconducting leads. Upper panel shows nearly momentum independent  $B_{\vec q}(\omega)$ (red dashed) and real part $A_{\vec q}(\omega)$ of inverse retarded propagator $[D_{12,\vec q}^R(\omega)]^{-1}$ for different momenta $\vec q$ with energy close to $\omega_{\vec q} \approx e V_0 - 2|\Delta|$. The photon decay rate $\eta$ is chosen to fulfill the laser threshold relation in Eq.~\eqref{lm4} in the superconducting state. The lasing regime is bounded by momenta $\vec q_1$ and $\vec q_2$ which correspond to the photon modes where $A_{\vec q}(\omega)$ and $B_{\vec q}(\omega)$ have simultaneous zeros (for some frequency $\omega$). Note the strong renormalization of the photon frequency close to $\omega = e V_0 - 2 |\Delta|$ which increases the lasing frequency window shown in the lower panel (in light blue).  
Lower panel shows divergence of luminescence for photon modes $\bfq_{1}$ and $\bfq_2$ and lasing regime (light blue) in between. }
\label{ans1}
\end{figure}

In Eqs.~\eqref{eq:20}-\eqref{prop} we have given the propagators within the RPA approximation. For normal conducting leads, the retarded function reduces to $[D_{12,\vec q}^R(\omega)]^{-1} = \omega-\omega_{\vec q} - \Pi_{21,\vec q}^R(\omega)$ and thus 
\begin{align}
\label{eq:28}
  A_{\vec q}(\omega) &= \omega-\omega_{\vec q}^* \\
  B_{\vec q}(\omega) &= - \Im \Pi_{21,\vec q}^{\text{el},R}(\omega)- \Im \Pi_{21,\vec q}^{\text{bath},R}(\omega)  \,.
\label{lm3}
 \end{align}  
Here, $\omega^*_\bfq = \omega_\bfq + \Re \Pi_{21,\vec q}^R$ denotes the renormalized photon frequency, where the real part of the self-energy is only weakly frequency dependent. The imaginary part $\Im \Pi_{21,\vec q}^{\text{el},R}(\omega)$ is given in Eq.~\eqref{nse1} and $\Im \Pi_{21,\vec q}^{\text{bath},R}(\omega) $ in Eq.~\eqref{pb2}. The bath contribution to the self-energy fulfills $\Im \Pi_{21,\vec q}^{\text{bath},R}(\omega)\approx -\eta <0$ for all frequencies $\omega$. In contrast, the electronic contribution $\Im \Pi_{21,\vec q}^{\text{el},R}(\omega)$ changes sign at $\omega = e V_0$ and is positive for frequencies below resonance $\omega < eV_0$ (see Fig.~\ref{NormalSE}). This follows from the electronic population inversion and describes photon production via transitions of conduction electrons to the valence band (see Fig.~\ref{bandsnc}). Depending on the size of the bath decay rate $\eta$ the total imaginary part $B_{\vec q}(\omega)$ may now be negative for frequencies $\omega < eV_0$. Photon excitations with a resonance energy $\omega_{\vec q}^*$ for which this is the case violate the analytical requirements of a retarded propagator. This indicates the breakdown of our assumption of a steady-state with a finite photon number. At this point, the system exhibits lasing~\cite{raey}. 

This breakdown can also be observed as a divergence of the luminescence $\mathcal L(\omega_{\vec q}) = \erw{b_{\vec q}^\dagger b_{\vec q}} = \frac{i}{2\pi}\intinfty d\omega D_{12,\vec q}^<(\omega)$, where
\begin{align}
\label{lm2c}
  D_{12,\vec q}^<(\omega) &= - \frac{i}{2} \frac{C_{\vec q}(\omega)-2B_{\vec q}(\omega)}{A_{\vec q}(\omega)^2+B_{\vec q}(\omega)^2}  \, . 
\end{align}
If the bath decay rate $\eta < \max |\Im \Pi_{21,\vec q}^{\text{el},R}(\omega)| \approx \pi \abs{g_0}^2 \rho_F$, the imaginary part vanishes $B_{\vec q}(\mu_{\vec q})=0$ for some frequency $\mu_{\vec q} = e V_0 - \eta v_F |\bfq|/(\pi \abs{g_0}^2 \rho_F)$. Around $\mu_{\vec q}$ one can linearize $B_{\vec q}(\omega) \approx \beta (\omega-\mu_{\vec q})$ with $\beta= \pi \abs{g_0}^2 \rho_F/(v_F |\bfq|)$. If the excitation energy $\omega^*_{\vec q} \approx \mu_{\vec q}$  is close to this critical energy, we can approximate the luminescence as
\begin{align}
\mathcal L(\omega^*_{\vec q})&= \frac{1}{2} \intinfty \frac{d\omega}{2\pi} \frac{C_{\vec q}(\omega)-2B_{\vec q}(\omega)}{A_{\vec q}(\omega)^2+B_{\vec q}(\omega)^2}   \nonumber \\
&\approx \frac{1}{2}\intinfty \frac{d\omega}{2\pi} \frac{C_{\vec q}(\mu_{\vec q})-2B_{\vec q}(\mu_{\vec q})}{(\omega-\omega^*_{\vec q})^2+\beta^2 (\omega-\mu_{\vec q})^2}     \nonumber \\
&= \frac{C_{\vec q}(\mu_{\vec q})}{4 \beta \abs{\omega^*_{\vec q}- \mu_{\vec q}}} \,.
\end{align}
We have used that the dominant part of the integral comes from the region around $\omega \approx \mu_{\vec q}, \omega^*_{\vec q}$. We note that the renormalization of the photon frequency is a small effect for normal conducting leads $\omega_\bfq \approx \omega_\bfq^*$. 

As shown in Fig.~\ref{ans0}, the photon number diverges like $\mathcal L(\omega^*_{\vec q}) \sim (\omega^*_{\vec q}- \mu_{\vec q})^{-1}$ from above, where the luminescence is finite. This behavior is reminiscent of Bose-Einstein condensation in an ideal gas. The divergence occurs for photons with momentum $\vec q_c$ where the zeros of the real part $A_{\vec q}(\omega)$ and of the imaginary part $B_{\vec q}(\omega)$ occur at the same frequency $\omega_{\bfq_c}^*$. Neglecting any frequency renormalization $\omega_\bfq = \omega^*_\bfq$, we find 
\begin{align}
  \label{eq:29}
  \abs{\vec q_c} = \frac{e V_0}{ c+ \eta \frac{v_F}{\pi \abs{g_0}^2 \rho_F}}
\end{align}
Due to its population inversion the electronic system induces an effective chemical potential $\mu_{\vec q_c}= c \abs{\vec q_c}$ for the photons. Within our approach, we cannot calculate the luminescence for frequencies $\omega_{\vec q} < \omega_{\vec q_c}$ since our assumption of a steady-state with a finite photon number breaks down. The LED produces more photons in these modes as the external bath can absorb. For these frequencies one needs to perform perturbation theory around a lasing state instead.

 To summarize, the system exhibits lasing if the imaginary part of the retarded self-energy becomes positive for positive frequencies. The luminescence diverges for that photon mode $\bfq_c$ for which both real and imaginary part of the retarded photon propagator exhibits zeros $A_{\vec q_c}(\omega) =B_{\vec q_c}(\omega) = 0$ for at least one $\omega$. 
\begin{figure}[t!]
\centering
\includegraphics[width=\linewidth]{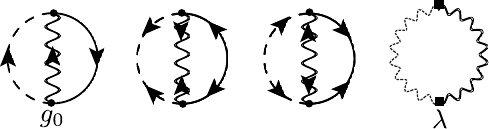}
\caption{Feynman graphs contributing to the rate equation of the photon number (see Eq.~\eqref{eq:30}). Solid (dashed) lines denote conduction (valence) electron propagators, solid double wiggly lines denote dressed photon propagators given in Eqs.~\eqref{eq:20}-~\eqref{prop} and dashed wiggly lines denote bath photon propagators. The first three diagrams represent the contribution to the rate from emission and absorption of photons involving electronic transitions in the the superconducting LED. The fourth graph describes the photon exchange with the external bath.}
\label{rate}
\end{figure}

Turning to the case of superconducting leads, we observe that the electronic contribution to the self-energy is strongly increased around $\omega  = e V_0 - 2 |\Delta|$ (see Fig.~\ref{NormalSE}). The laser threshold thus corresponds to a larger value of the bath absorption rate

\begin{align}
\eta_c = \begin{cases} \pi \abs{g_0}^2 \rho_F \,, &\text{normal conducting leads} \\ 
  \pi \abs{g_0}^2 \rho_F  \frac{\pi |\Delta|}{2 v_F |\bfq|}\,, &\text{superconducting leads}\label{lm4} \end{cases}   \,,
\end{align}
where $\pi|\Delta|/(2 v_F |\bfq|) > 1$ for our realistic choice of parameters (see Sec.~\ref{sec:energy-scales-system}). In Fig.~\ref{ans1} we show $A_{\vec q}(\omega),B_{\vec q}(\omega)$ and the luminescence $\mathcal{L}(\omega^*_\bfq)$ for superconducting leads for $\pi \abs{g_0}^2 \rho_F < \eta <\frac{\pi^2 \abs{g_0}^2 \rho_F |\Delta|}{2 v_F |\bfq|} $ such that there is no lasing in the absence of superconductivity. As the simultaneous zeros in the upper panel of the figure indicate, superconductivity leads to lasing in the frequency range $\omega_{\vec q_1} < \omega_{\vec q} < \omega_{\vec q_2} \approx e V_0 - 2 |\Delta|$. The luminescence clearly diverges as $1/(\omega_\bfq - \omega_{\bfq_{1,2}})$ at the border of this region. Within this region one needs to perform perturbation theory around a lasing state, which is beyond the scope of this work. 

To conclude, superconductivity leads to a sharp increase of the photon production rate of the LED around the frequency $\omega_\bfq \approx e V_0 - 2 |\Delta|$. This induces a sharp increase of the electroluminescence $\mathcal{L}(\omega_\bfq)$ and a smaller laser threshold $\eta_c$. 
\subsection{Lasing conditions from analyzing rate equations}
\label{sec:rate-equations}
The lasing condition in Eq.~\eqref{lm4} can also be derived using rate equations. This approach is appealing due to its simplicity and physical transparency. It begins with calculating the change of the number of bosons $\hat n_b = \sum_{\vec q} b_{\vec q}^\dagger b_{\vec q}$, which is given by
\begin{equation}
\Gamma = \frac{d}{dt} \av{\hat{n}_b (t)} = - i \av{ [\hat{n}_b(t), \hat{H}(t)] } \,,
 \label{rp1}
\end{equation}
where the time-dependent operators $\hat n_b(t), \hat H(t)$ are defined in the Heisenberg picture. The expectation value $\av{\mathcal{O}}$ of an operator $\mathcal{O}$ is defined with respect to the full Hamiltonian $H$ in Eq.~\eqref{mod1}. The commutator can be evaluated to 

\begin{align}
\label{eq:30}
\bigl[\hat n_b(t), \hat H(t) \bigr] &= \sum_{\vec k,\vec k', \sigma} \bigl[\bar{g}(t) v^\dag_{\vec k,\sigma}(t) b^\dag_{\vec k'-\vec k}(t) c_{\vec k',\sigma}(t)  \\
& \quad - g(t) v_{\vec k,\sigma}(t) b_{\vec k'-\vec k}(t) c^\dag_{\vec k',\sigma}(t) \bigr] \nonumber \\
& \quad + \sum_{\vec p,\vec q}  \bigl[ \lambda_{\vec p,\vec q} b^\dag_{\vec p}(t)  a_{\vec q}(t)- \bar \lambda_{\vec q,\vec p}  a^\dag_{\vec p}(t)  b_{\vec q}(t) \bigr] \nonumber  \, .
\end{align}
The expectation value of this commutator can be conveniently evaluated within perturbation theory in $g_0$ using the Keldysh formalism~\cite{TikhodeevUeba-OxfordUnivPress-2008}. The total photon production rate $\Gamma =\Gamma_ {\text{SLED}} + \Gamma_ {\text{bath}} $ is composed of two contributions: first, $\Gamma_{\text{SLED}}$ which describes the emission and absorption of photons due to electronic transitions in the LED and second, $\Gamma_{\text{bath}}$, which characterizes the photon exchange with the external bath. An explicit calculation, which takes the diagrams shown in Fig.~\ref{rate} into account, yields
\begin{widetext}
\begin{align}
\label{eq:31}
\Gamma_ {\text{bath}} &=  - 2 \sum_{\vec q} \int \frac{d\omega}{\pi} \Im \Pi_{\vec q}^{\text{bath},R}(\omega)  \Im D_{12,\vec q}^A(\omega) \bigl[ n^{\text{bath}}(\omega) -n_{\vec q}(\omega) \bigr] \\
\label{rp3}
\Gamma_ {\text{SLED}} &=  2 \sum_{\vec q} \int \frac{d\omega}{\pi} \Im \Pi_{21,\vec q}^{\text{el},R}(\omega) \Im D_{12,\vec q}^A(\omega) \biggl( \bigl[1+n_{\vec q}(\omega)\bigr] \theta(eV_0-\omega) + n_{\vec q}(\omega) \theta(\omega-eV_0) \biggr)  \nonumber \\
& \qquad - \frac{1}{\pi} \sum_{\vec q} \int \frac{d\omega}{\pi} \Re \bigl[  \tilde \Pi_{11,\vec q}^{\text{el},R}(\omega) D_{11,\vec q}^A(\omega)  \bigr] \, .
\end{align}  
Here, $n_{\vec q}(\omega)$ denotes the photon distribution function in the (superconducting) LED system and $n^{\text{bath}}(\omega)$ denotes the photon distribution in the external bath. For a thermal bath, $n^{\text{bath}}(\omega)$ is given by the Bose-Einstein distribution function. The coupling to the bath drives the photon distribution in the system $n_\bfq$ towards the external bath distribution $n_{\vec q}(\omega) \rightarrow n^{\text{bath}}(\omega)$. In contrast, the LED produces photons via spontaneous as well as stimulated emission for $\omega<e V_0$ and absorbs them for $\omega > e V_0$.

We evaluate the integrals in Eq.~\eqref{eq:31} and~\eqref{rp3} to leading order in the electron-photon coupling $g_0$. To this order, we only get an on-shell contribution $\Im D_{12,\vec q}^A(\omega) = \pi \delta(\omega-\omega_{\vec q})$, because higher order terms in $\Im \tilde D_{11,\vec q}^A(\omega) \sim g_0^2$ are neglected as the self-energies are already of order $\mathcal{O}(g_0^2)$. As a result, the frequency integration enforces $n_{\vec q}(\omega) = n(\omega_{\vec q})$ and the total photon production rate simplifies to
\begin{align}
\Gamma &=  2 \sum_{\vec q}   \biggl[ -\Im  \Pi_{\vec q}^{\text{bath},R}(\omega_{\vec q})  \bigl[ n^{\text{bath}}(\omega_{\vec q}) -n(\omega_{\vec q}) \bigr]  + \Im  \Pi_{21,\vec q}^{\text{el},R}(\omega_{\vec q})   \biggl( \bigl[1+n(\omega_{\vec q})\bigr] \theta(eV_0-\omega_{\vec q}) + n(\omega_{\vec q}) \theta(\omega_{\vec q}-eV_0) \biggr)    \biggr]   
\end{align}
In a steady state the photon emission and absorption are balanced and the total rate $\Gamma = \sum_{\vec q} \Gamma_{\vec q}$ is zero for all photon modes $\vec q$. Employing that $\Im \Pi_{\vec q}^{\text{bath},R}(\omega_{\vec q}) < 0$ and $\sign \bigl[\Im\Pi_{21,\vec q}^{\text{el},R}(\omega_{\vec q}) \bigr] \sim \sign(eV_0 - \omega_{\vec q})$ we can infer the photon distribution in the steady state as
\begin{align}
\label{eq:17}
 n(\omega_{\vec q})       = \frac{ \abs{\Im \Pi_{21,\vec q}^{\text{el},R}(\omega_{\vec q}) } \theta(e V_0-\omega_{\vec q}) + \abs{\Im \Pi_{\vec q}^{\text{bath},R}(\omega_{\vec q})} n^{\text{bath}}(\omega_{\vec q}) }{-\Im \Pi_{\vec q}^{\text{bath},R}(\omega_{\vec q}) -\Im \Pi_{21,\vec q}^{\text{el},R}(\omega_{\vec q}) } \,.
 \end{align}  
 \end{widetext}
For $\omega_{\vec q} > eV_0$ the distribution function is positive and finite. For $\omega_{\vec q} < eV_0$, however, the imaginary part of the total self-energy $\Im \Pi_{21,\vec q}^{R}(\omega_{\vec q}) = \Im \Pi_{\vec q}^{\text{bath},R}(\omega_{\vec q}) +\Im \Pi_{21,\vec q}^{\text{el},R}(\omega_{\vec q})$ in the denominator of Eq.~\eqref{eq:17} may become zero. At this point, the production rate of the LED is exactly equal in magnitude to the decay rate into the bath. This results in the divergence of the photon distribution for energies $\omega_{\vec q}$ for which $\Im \Pi_{21,\vec q}^{R}(\omega_{\vec q})=0$. It follows that the laser threshold in the (superconducting) LED is given by our previous result in Eq.~\zref{lm4}.

\section{Conclusions}
\label{sec:conclusions-outlook}
We have investigated the electroluminescence and photonic properties of a forward biased $p$-$n$ junction in proximity to superconducting leads. We have shown that superconductivity leads to a significant enhancement of the luminescence of the light-emitting diode in a frequency window of the order of $(v_F e V_0)^2/(c^2 |\Delta|) \approx 10^{-3} eV$. This effect stems from the increased density of states at the edges of the superconducting gap in the electronic bandstructure of the valence and conduction bands. The increased photon production rate also reduces the lasing threshold in the system. By summing the complete infinite order RPA perturbation series in the photon-electron coupling, we were able to show that an additional luminescence peak occurs on resonance due to the tunneling of Cooper pairs from the conduction to the valence band. In addition, such a superconducting light-emitting diode emits two-mode squeezed light. The squeezing angle is controlled by the superconducting phase difference $\phi_c-\phi_v$ between conduction and valence band. This proves that one may transfer the macroscopic coherence of an electronic superconducting condensate to photon pairs and manipulate the photonic coherence electronically. 
 
\acknowledgments
We acknowledge useful discussions with P. Baireuther, U. Briskot, B. Narozhny, M. Sch\"utt and J. Schmalian. The Young Investigator Group of P.P.O. received
financial support from the “Concept for the Future” of the Karlsruhe Institute of Technology (KIT), Germany, within the framework of the German Excellence Initiative. 
\appendix
\section{Effective photon action}   
\label{appepa}
The unperturbed fermionic and bosonic part of the action in Eq.~\eqref{eq:1} can be written in the basis of the R,A,K fields as
\begin{align}
\label{eq:33}
&S_{c}+S_{v} = \sum_{\vec k} \intinfty dt dt' \, \hat \Psi_{\vec k}^\dagger(t) \hat G_{0,\vec k}^{-1}(t,t') \hat \Psi_{\vec k}(t') \, ,  \\
\label{epa1}
&S_{ph} = \frac{1}{2} \sum_{\vec q}\intinfty dt dt' \, \hat \Phi_{\vec q}^T(t) \hat D_{0,\vec q}^{-1}(t,t') \hat \Phi_{-\vec q}(t') \, .  
\end{align}
where we defined the fermionic and bosonic spinors and matrix Green's functions in Section~\ref{basmodsec}. The electron-photon interaction and photon-bath interaction can also be written in terms of these spinors. 

\begin{widetext}
\subsection{Integrating out the electrons} 
\label{sec:integr-out-electr}
Let us consider first the electron-photon coupling
\begin{align}
S_{int} &= \sum_{\vec k,\vec k'\atop \sigma} \int_{\mathcal C} dt   \biggl( g(t)  b_{\vec k-\vec k'}(t) \bar c_{\vec k,\sigma}(t) v_{\vec k',\sigma}(t) + \text{h.c.}   \biggr) = \sum_{\vec k,\vec k'}  \int_{\mathcal C} dt \De \Psi_{\vec k}^\dagger(t) \bigl[\sum_{i=1,2} \sqrt{2} \De g_i(t) \ring \Phi_{\vec k-\vec k',i}(t)\bigr] \De \Psi_{\vec k'}(t)   \label{epa2}
\end{align}
where $\ring \Phi_{\vec k} = (\ring \Phi_{\vec k,1}, \ring \Phi_{\vec k,2})^T=(b_{\vec k}, \bar b_{-\vec k})^T$ and we define the matrices in the extended Nambu-space as
\begin{align}
\De g_1(t)  = \frac{1}{\sqrt{2}}\mx{ 0 &0& 0 & 0 \\ g(t) & 0 & 0 & 0 \\ 0 & 0 & 0 & -g(t) \\ 0 & 0 & 0 & 0}     \platz  \De g_2(t)  =\frac{1}{\sqrt{2}} \mx{ 0 &\bar g(t) & 0 & 0 \\ 0 & 0 & 0 & 0 \\ 0 & 0 & 0 & 0 \\ 0 & 0 & -\bar g(t) & 0}   \label{epa3}
\end{align}
The Keldysh contour can now be expressed by two integrations on the real axis for the $+$ and $-$ fields
\begin{align}
S_{int} &=\sum_{\vec k,\vec k'}  \intinfty dt \,  \sum_{\alpha=\pm} \alpha \cdot \bigl[\De \Psi_{\vec k}^\alpha(t)\bigr]^\dagger \bigl[\sum_{i=1,2} \sqrt{2} \De g_i(t)  \Phi_{\vec k-\vec k',i}^\alpha (t)\bigr] \De \Psi_{\vec k'}^\alpha (t)   \nonumber \\
&= \sum_{\vec k,\vec k'}  \intinfty dt \,   \hat \Psi_{\vec k}^\dagger(t) \underbrace{\bigl[\sum_{\alpha=\q, \cl \atop i=1,2} \hat \gamma^\alpha \De g_i(t)  \Phi_{\vec k-\vec k',i}^\alpha (t)\bigr]}_{\hat V_{\vec k,\vec k'}(t)} \hat \Psi_{\vec k'} (t)   \label{epa4}
\end{align}
and finally we have rewritten the action in the R,A,K basis using $\hat \gamma^{\text{cl}} = \hat{\openone}$ and $\hat \gamma^{\text{q}} = \hat{\sigma}_x$ in Keldysh space. Here, we use the convention that we always first evaluate the Keldysh matrix structure and after that the structure of the fermionic Nambu-space $\Delta$ or bosonic particle hole space $\circ$. Now, we can integrate out the fermions easily as we have a quadratic action
\begin{align}
\int D[\Psi,\bar \Psi] e^{i S_{ph}} e^{i [S_c+S_v+S_{\text{int}}]} &= e^{i S_{ph}} \int D[\Psi,\bar \Psi] e^{i \sum_{\vec k,\vec k'} \intinfty dt dt'\,  \hat \Psi_{\vec k}^\dagger(t) \biggl( \hat G_{0,\vec k}^{-1}(t,t') \delta_{\vec k,\vec k'}+  \hat V_{\vec k,\vec k'}(t) \delta(t-t')    \biggr) \hat \Psi_{\vec k}(t')   } \nonumber \\
&= e^{i S_{ph}+ \tr \ln \bigl[ 1+ \hat G_0 \hat V]} = e^{i S_{ph,\text{eff}}^{\text{el}}}   \label{epa5}
\end{align}
Expanding the trace log we find in leading order in $g$ that
\begin{align}
\tr \ln \bigl[ 1+ \hat G_0 \hat V] &= \sum_{n=1}^\infty \frac{(-1)^{n+1}}{n} \tr\bigl[ (\hat G_0 \hat V)^n \bigr] = - \frac{1}{2}\tr\bigl[ (\hat G_0 \hat V)^2 \bigr] + \ldots  \nonumber \\
&\approx  - \frac{1}{2} \int dt dt' \sum_{\vec k, \vec k'} \tr \bigl[ \hat G_{0,\vec k}(t',t)  \hat V_{\vec k,\vec k'}(t)   \hat G_{\vec k'}(t,t')   \hat V_{\vec k',\vec k}(t')  \bigr]    \nonumber \\
&= - \frac{1}{2} \int dt dt' \sum_{\vec k, \vec q} \tr \biggl[ \hat G_{0,\vec k}(t',t) \bigl[\sum_{\alpha=\q, \cl \atop i=1,2} \hat \gamma^\alpha \De g_i(t)  \Phi_{\vec q,i}^\alpha (t)\bigr]   \hat G_{0,\vec k-\vec q}(t,t')  \bigl[\sum_{\beta=\q, \cl \atop j=1,2} \hat \gamma^\beta \De g_j(t')  \Phi_{-\vec q,j}^\beta (t')\bigr]   \biggr]  \nonumber \\
&=  - \frac{1}{2} \int dt dt' \sum_{\vec q}  \sum_{i,j=1,2 \atop \alpha,\beta = \q, \cl} \Phi_{\vec q,i}^\alpha (t) \underbrace{\biggl( \sum_{\vec k} \tr \bigl[ \hat G_{0,\vec k}(t',t)  \hat \gamma^\alpha \De g_i(t)     \hat G_{0,\vec k-\vec q}(t,t')  \hat \gamma^\beta \De g_j(t')  \bigr]  \biggr)  }_{:= i [\Pi^{\text{el}}]_{ij,\vec q}^{\alpha \beta}(t,t')}  \Phi_{-\vec q,j}^\beta(t')    \nonumber \\
&=  - \frac{i}{2} \int dt dt' \sum_{\vec q}  \hat \Phi_{\vec q}^T(t) \hat \Pi_{\vec q}^{\text{el}}(t,t') \hat \Phi_{-\vec q,j}(t')   \label{epa6}
\end{align}
where we still have to evaluate the trace over the Keldysh and Nambu structure. We  get 
\begin{align}
S_{ph,\text{eff}}^{\text{el}} = \frac{1}{2} \sum_{\vec q}\intinfty dt dt' \, \hat \Phi_{\vec q}^T(t)  \biggl[\hat D_{0,\vec q}^{-1}(t,t')  -    \hat \Pi_{\vec q}^{\text{el}}(t,t')  \biggr] \hat \Phi_{-\vec q}(t')   \label{epa7}
\end{align}
with
\begin{align}
\hat \Pi_{\vec q}^{\text{el}}(t,t') &= \mx{ 0 &  \ring \Pi_{\vec q}^{\text{el},A}(t,t') \\ \ring \Pi_{\vec q}^{\text{el},R}(t,t')  & \ring \Pi_{\vec q}^{\text{el},K}(t,t') }   \nonumber \\
\ring \Pi_{\vec q}^{\text{el},R}(t,t') &= \mx{ \bigl[ \Pi^{\text{el}} \bigr]^{\q, \cl}_{11,\vec q} (t,t') & \bigl[ \Pi^{\text{el}} \bigr]^{\q, \cl}_{12,\vec q} (t,t')  \\ \bigl[ \Pi^{\text{el}} \bigr]^{\q, \cl}_{21,\vec q} (t,t') & \bigl[ \Pi^{\text{el}} \bigr]^{\q, \cl}_{22,\vec q} (t,t')   }  \nonumber \\
\ring \Pi_{\vec q}^{\text{el},A}(t,t') &= \mx{ \bigl[ \Pi^{\text{el}} \bigr]^{\cl,\q}_{11,\vec q} (t,t') & \bigl[ \Pi^{\text{el}} \bigr]^{\cl,\q}_{12,\vec q} (t,t')  \\ \bigl[ \Pi^{\text{el}} \bigr]^{\cl,\q}_{21,\vec q} (t,t') & \bigl[ \Pi^{\text{el}} \bigr]^{\cl,\q}_{22,\vec q} (t,t')   }   \nonumber \\
\ring \Pi_{\vec q}^{\text{el},A}(t,t') &= \mx{ \bigl[ \Pi^{\text{el}} \bigr]^{\cl,\cl}_{11,\vec q} (t,t') & \bigl[ \Pi^{\text{el}} \bigr]^{\cl,\cl}_{12,\vec q} (t,t')  \\ \bigl[ \Pi^{\text{el}} \bigr]^{\cl,\cl}_{21,\vec q} (t,t') & \bigl[ \Pi^{\text{el}} \bigr]^{\cl,\cl}_{22,\vec q} (t,t')   }   \nonumber \\
[\Pi^{\text{el}}]_{ij,\vec q}^{\alpha \beta}(t,t') &= - i \sum_{\vec k} \tr \bigl[  \hat \gamma^\alpha \De g_i(t)     \hat G_{0,\vec k}(t,t')  \hat \gamma^\beta \De g_j(t') \hat G_{0,\vec k+\vec q}(t',t)  \bigr]       \label{epa8}
\end{align}   

\subsection{Integrating out the photon bath}   \label{couplingtophotonpath}
Following the same lines as the previous calculation, we want to integrate out the bath photons in \zref{mod3.1} to find the corresponding self-energy for the SLED photons, see also Ref.~[\onlinecite{PhysRevLett.96.230602,PhysRevB.75.195331}]. It is straightforward to show that the corresponding self-energy is then given by
\begin{align}
\hat \Pi_{\vec q}^{\text{bath}}(t,t') &= \mx{ 0 &  \ring \Pi_{\vec q}^{\text{bath},A}(t,t') \\ \ring \Pi_{\vec q}^{\text{bath},R}(t,t')  & \ring \Pi_{\vec q}^{\text{bath},K}(t,t') }   \nonumber \\
\hat \Pi_{\vec q}^{\text{bath},R/A/K}(t,t') &=  \sum_{\vec p} \abs{\lambda_{\vec q,\vec p}}^2 \mx{0 & d_{\text{bath},\vec p}^{A/R/K}(t',t)  \\   d_{\text{bath},\vec p}^{R/A/K}(t,t')&0} \nonumber \\
d_{\text{bath},\vec q}^{R/A}(\omega) &= \frac{1}{\omega-\omega_{\vec q} \pm i 0}   \nonumber \\
d_{\text{bath},\vec q}^{K}(\omega) &= B_0(\omega) \bigl[   d_{\text{bath},\vec q}^{R}(\omega) -d_{\text{bath},\vec q}^{A}(\omega)\bigr]
\end{align}
The assumed (SLED photon) momentum-independent coupling $\lambda_{\vec p,\vec p'} = \lambda(\omega_{\vec p'})$ then leads to the simple relation (we omit the unimportant since featureless real part)
\begin{align}
\hat \Pi_{\vec q}^{\text{bath},R/A}(\omega) &= \mx{ 0 & \pm i \eta(-\omega) \\ \mp i \eta(\omega) & 0}   \\
\hat \Pi_{\vec q}^{\text{bath},K}(\omega) &= B_0(\omega) \bigl[   \hat \Pi_{\vec q}^{\text{bath},R}(\omega)-\hat \Pi_{\vec q}^{\text{bath},A}(\omega)\bigr]
\end{align}
with
\begin{align}
\eta(\omega) &= \Im \biggl[ \sum_{\vec p}\abs{\lambda(\omega_{\vec p})}^2  d_{\text{bath},\vec p}^{A}(\omega)   \biggr] = \pi \sum_{\vec p} \abs{\lambda(\nu_{\vec p})}^2 \delta(\omega-\omega_{\vec p})  = \pi \int d\epsilon \,  N_{\text{bath}}(\epsilon) \abs{\lambda(\epsilon)}^2 \delta(\omega-\epsilon)    \nonumber \\
&=\pi \abs{\lambda(\omega)}^2 N_{\text{bath}}(\omega)
\end{align}
where $N_{\text{bath}} = \sum_{\vec p} \delta(\omega-\nu_{\vec p})$ is the DOS of the photon bath. We choose an Markovian (or Ohmic) bath, where the DOS and the coupling are constant in frequency for all $\omega \gtrsim \Lambda$ (where $\Lambda \ll eV_0$) and thus give rise to a constant absorption rate of the bath for the photons produced by the SLED (with characteristic energy $\omega_{\vec q} \approx eV_0$). We use the parametrization
\begin{align}
\eta(\omega) = \eta \cdot \frac{\omega^2}{\omega^2+\Lambda^2} \, \theta(\omega)     
\end{align}
with $\eta(0)=0$ (which is important as explained in the main text). The derivation from the constant behavior for small $\omega \ll e V_0$ does not change the results and features of the SLED presented in this paper as we are interested at photons with $\omega_{\vec q} \approx e V_0$. 

The total effective action then reads
\begin{align}
S_{ph,\text{eff}} = \frac{1}{2} \sum_{\vec q}\intinfty dt dt' \, \hat \Phi_{\vec q}^T(t)  \biggl[\hat D_{0,\vec q}^{-1}(t,t')  -    \hat \Pi_{\vec q}^{\text{el}}(t,t') -    \hat \Pi_{\vec q}^{\text{bath}}(t,t') \biggr] \hat \Phi_{-\vec q}(t')   \label{epa10}
\end{align}

\section{Calculation of the self-energy $\hat \Pi^{\text{el}}$}   \label{appelse}
We already derived the expressions for the bosonic self-energy caused by the superconducting electrons in the system in \zref{epa8}. Let us now first derive the causal structure of this self-energy by performing the trace over the Keldysh indices
\begin{align}
[\Pi^{\text{el}}]_{ij,\vec q}^{R}(t,t') &=  \bigl[ \Pi^{\text{el}} \bigr]^{\q, \cl}_{ij,\vec q} (t,t') = - i \sum_{\vec k} \tr \bigl[  \hat \gamma^\q \De g_i(t)     \hat G_{0,\vec k}(t,t')  \hat \gamma^\cl \De g_j(t') \hat G_{\vec k+\vec q}(t',t)  \bigr]   \nonumber \\
&\verweis{mod6} - i \sum_{\vec k} \tr_\Delta \tr_{\wedge} \biggl[  \mx{ 0 & 1\\ 1 & 0} \De g_i(t)     \mx{ \De  G_{0,\vec k}^R(t,t') & \De  G_{0,\vec k}^K(t,t') \\ 0 & \De  G_{0,\vec k}^A(t,t')}  \mx{ 1 & 0 \\ 0 &1} \De g_j(t') \mx{ \De  G_{0,\vec k+\vec q}^R(t',t) & \De  G_{0,\vec k+\vec q}^K(t',t) \\ 0 & \De  G_{0,\vec k+\vec q}^A(t',t)} \biggr]   \nonumber \\
&= - i \sum_{\vec k} \tr_\Delta \bigl[ \De g_i(t) \De  G_{0,\vec k}^R(t,t')     \De g_j(t')   \De  G_{0,\vec k+\vec q}^K(t',t)  +  \De g_i(t) \De  G_{0,\vec k}^K(t,t')     \De g_j(t')   \De  G_{0,\vec k+\vec q}^A(t',t) \bigr]   \label{cse1}   \\
[\Pi^{\text{el}}]_{ij,\vec q}^{A}(t,t') &= - i \sum_{\vec k} \tr_\Delta \bigl[ \De g_i(t) \De  G_{0,\vec k}^A(t,t')     \De g_j(t')   \De  G_{0,\vec k+\vec q}^K(t',t)  +  \De g_i(t) \De  G_{0,\vec k}^K(t,t')     \De g_j(t')   \De  G_{0,\vec k+\vec q}^R(t',t) \bigr] \label{cse2}    \\
[\Pi^{\text{el}}]_{ij,\vec q}^{A}(t,t') &= - i \sum_{\vec k} \tr_\Delta \biggl[ \De g_i(t) \De  G_{0,\vec k}^K(t,t')     \De g_j(t')   \De  G_{0,\vec k+\vec q}^K(t',t)    \nonumber \\
& \hspace{2cm}-  \De g_i(t) \bigl[\De  G_{0,\vec k}^R(t,t') - \De  G_{0,\vec k}^A(t,t')\bigr]    \De g_j(t')  \bigl[ \De  G_{0,\vec k+\vec q}^R(t',t)- \De  G_{0,\vec k+\vec q}^A(t',t)\bigr]  \biggr]  \label{cse3}   
\end{align}
which are just the usual forms of the retarded, advanced and Keldysh self-energies as known from standard text books. The next step will be the evaluation of the Nambu trace. Let us first write define the \textit{absolute} anomalous propagator ($\alpha=R,A,K$)
\begin{align}
\begin{split}
P_{0,\vec k,v/c}^\alpha(t,t') &= e^{- i \phi_{v/c}} \abs{P}_{0,\vec k,v/c}^\alpha(t-t')  \\ 
\bar P_{0,\vec k,v/c}^\alpha(t,t') &= e^{ i \phi_{v/c}} \abs{P}_{0,\vec k,v/c}^\alpha(t-t')   \\
\abs{P}_{0,\vec k,v/c}^{R/A}(\omega) &= - \frac{\abs{\Delta_{c/v}}}{(\omega\pm i0)^2 - \xi_{v/c}(\vec k)^2- \abs{\Delta_{c/v}}^2} 
\end{split} \label{cse4} 
\end{align}
where we separated the phase of the superconducting gaps $\Delta_{v/c}= e^{i \phi_{v/c}} \abs{\Delta_{v/c}}$. Defining also the phase of the electron-photon coupling $g_0=e^{i \phi_g} \abs{g_0}$, the absolute time $T=(t+t')/2$ and relative time $\tau=t-t'$ we can evaluate
\begin{align}
\begin{split}
\tr_\Delta \bigl[\De g_1(t) \De  G_{0,\vec k}^\alpha(t,t')     \De g_1(t')   \De  G_{0,\vec k+\vec q}^\beta(t',t)  \bigr] &= -\frac{g(t) g(t')}{2} \bigl[ P_{0,\vec k,v}^\alpha(t,t') \bar P_{0,\vec k+\vec q,c}^\beta(t',t) + \bar P_{0,\vec k,c}^\alpha(t,t')  P_{0,\vec k+\vec q,v}^\beta(t',t)      \bigr]     \\
&= - e^{i \phi(2T)} \abs{g_0}^2  \bigl[  \abs{P}_{0,\vec k,v}^\alpha(\tau) \abs{ P}_{0,\vec k+\vec q,c}^\beta(-\tau) +\abs{P}_{0,\vec k,c}^\alpha(\tau)  \abs{P}_{0,\vec k+\vec q,v}^\beta(-\tau)  \bigr]   \\ 
\tr_\Delta \bigl[\De g_1(t) \De  G_{0,\vec k}^\alpha(t,t')     \De g_2(t')   \De  G_{0,\vec k+\vec q}^\beta(t',t)  \bigr] &= \frac{g(t) \bar g(t')}{2} \bigl[ G_{0,\vec k,v}^{(p),\alpha}(t,t') G_{0,\vec k+\vec q,c}^{(p),\beta}(t',t) + G_{0,\vec k,c}^{(h),\alpha}(t,t')  G_{0,\vec k+\vec q,v}^{(h),\beta}(t',t)      \bigr]   \\
&=  e^{i e V_0 \tau}  \abs{g_0}^2 \bigl[ G_{0,\vec k,v}^{(p),\alpha}(\tau) G_{0,\vec k+\vec q,c}^{(p),\beta}(-\tau) + G_{0,\vec k,c}^{(h),\alpha}(\tau)  G_{0,\vec k+\vec q,v}^{(h),\beta}(-\tau)      \bigr]  \\
\tr_\Delta \bigl[\De g_2(t) \De  G_{0,\vec k}^\alpha(t,t')     \De g_1(t')   \De  G_{0,\vec k+\vec q}^\beta(t',t)  \bigr] &= \frac{\bar g(t)  g(t')}{2} \bigl[ G_{0,\vec k,v}^{(h),\alpha}(t,t') G_{0,\vec k+\vec q,c}^{(h),\beta}(t',t) + G_{0,\vec k,c}^{(p),\alpha}(t,t')  G_{0,\vec k+\vec q,v}^{(p),\beta}(t',t)      \bigr]  \\
&=  e^{-i e V_0 \tau}  \abs{g_0}^2 \bigl[ G_{0,\vec k,v}^{(h),\alpha}(\tau) G_{0,\vec k+\vec q,c}^{(h),\beta}(-\tau) + G_{0,\vec k,c}^{(p),\alpha}(\tau)  G_{0,\vec k+\vec q,v}^{(p),\beta}(-\tau)      \bigr] \\
\tr_\Delta \bigl[\De g_2(t) \De  G_{0,\vec k}^\alpha(t,t')     \De g_2(t')   \De  G_{0,\vec k+\vec q}^\beta(t',t)  \bigr] &= -\frac{\bar g(t) \bar g(t')}{2} \bigl[ \bar P_{0,\vec k,v}^\alpha(t,t')  P_{0,\vec k+\vec q,c}^\beta(t',t) +  P_{0,\vec k,c}^\alpha(t,t')  \bar P_{0,\vec k+\vec q,v}^\beta(t',t)      \bigr] \\
&= - e^{-i \phi(2T)} \abs{g_0}^2  \bigl[  \abs{P}_{0,\vec k,v}^\alpha(\tau) \abs{ P}_{0,\vec k+\vec q,c}^\beta(-\tau) +\abs{P}_{0,\vec k,c}^\alpha(\tau)  \abs{P}_{0,\vec k+\vec q,v}^\beta(-\tau)  \bigr]  
\end{split}  \label{cse5} 
\end{align}
with the rotating phase
\begin{align}
\phi(2T) =2 e V_0 T+  [\Delta_v \Delta_c^* g^2] = 2 e V_0 T -\phi_c+\phi_v+2 \phi_g    \label{cse6}
\end{align}
From \zref{epa8} and \zref{cse5} we see that we indeed find the structure presented in Eq.~\zref{ef1} in the photon particle-hole space $\circ$. Performing the Wigner transformation $f(\omega,T) =\mathcal W \bigl\{ f(\tau,T)  \bigr\}_{\tau,\omega} = \intinfty d\tau f(\tau,T) e^{i \omega \tau}$, which is just a Fourier transformation in relative time $\tau$, we have the identity
\begin{align}
\mathcal W \bigl\{e^{i a T} e^{i b \tau} A(\tau) B(-\tau)  \bigr\}_{\tau,\omega} = e^{i a T} \intinfty\frac{d\omega_1}{2\pi}  A(\omega_1) B(\omega_1-\omega-b)    \label{cse7}
\end{align}
In total, we can write down the photon self-energy part from the coupling to the electrons as
\begin{align}
\ring \Pi_{\vec q}^{\text{el},R/A}(\omega,T) &= \mx{ e^{i \phi(2T)} \tilde \Pi_{11,\vec q}^{\text{el},R/A}(\omega)  & \Pi_{12,\vec q}^{\text{el},R/A}(\omega) \\ \Pi_{21,\vec q}^{\text{el},R/A}(\omega) & e^{-i \phi(2T)} \tilde\Pi_{22,\vec q}^{\text{el},R/A}(\omega)}     \label{cse8} 
\end{align}
with
\begin{align}
\label{eq:34}
\tilde \Pi_{11,\vec q}^{\text{el},R/A}(\omega) &= i \abs{g_0^2} \sum_{\vec k} \intinfty \frac{d\omega_1}{2\pi}   \biggl[  \abs{P}_{0,\vec k,v}^{R/A}(\omega_1) \abs{ P}_{0,\vec k+\vec q,c}^K(\omega_1-\omega) +\abs{P}_{0,\vec k,v}^K(\omega_1) \abs{ P}_{0,\vec k+\vec q,c}^{A/R}(\omega_1-\omega)  \biggr]  \\
\label{eq:35}
\Pi_{12,\vec q}^{\text{el},R/A}(\omega) &= - i \abs{g_0^2} \sum_{\vec k} \intinfty \frac{d\omega_1}{2\pi}   \biggl[  G_{0,\vec k,v}^{(p),R/A}(\omega_1) G_{0,\vec k+\vec q,c}^{(p),K}(\omega_1-\omega_+) +G_{0,\vec k,v}^{(p),K}(\omega_1) G_{0,\vec k+\vec q,c}^{(p),A/R}(\omega_1-\omega_+) \biggr] \\
\label{eq:36}
\Pi_{21,\vec q}^{\text{el},R/A}(\omega) &= - i \abs{g_0^2} \sum_{\vec k} \intinfty \frac{d\omega_1}{2\pi}   \biggl[  G_{0,\vec k,v}^{(h),R/A}(\omega_1) G_{0,\vec k+\vec q,c}^{(h),K}(\omega_1-\omega_-) +G_{0,\vec k,v}^{(h),K}(\omega_1) G_{0,\vec k+\vec q,c}^{(h),A/R}(\omega_1-\omega_-) \biggr]   \\
\label{cse9} 
\tilde \Pi_{22,\vec q}^{\text{el},R/A}(\omega)  &= i \abs{g_0^2} \sum_{\vec k} \intinfty \frac{d\omega_1}{2\pi}   \biggl[  \abs{P}_{0,\vec k,v}^{R/A}(\omega_1) \abs{ P}_{0,\vec k+\vec q,c}^K(\omega_1-\omega) +\abs{P}_{0,\vec k,v}^K(\omega_1) \abs{ P}_{0,\vec k+\vec q,c}^{A/R}(\omega_1-\omega)  \biggr]
\end{align}
where we defined $\omega_\pm=\omega \pm e V_0$ and used several symmetry properties
\begin{align}
\label{eq:37}
 G_{0,\vec k,v/c}^{(p),R/A/K}(\omega) &= - G_{0,\vec k,v/c}^{(h),A/R/K}(-\omega)    \\
 \abs{P}_{0,k,v/c}^{R/A/K}(\omega) &=  \abs{P}_{0,k,v/c}^{A/R/K}(-\omega)  
\label{cse10}  
 \end{align} 
 of the fermionic propagators.  The corresponding Feynman graphs are presented in Fig.~\ref{Seel1}. Using the symmetries \zref{cse9} it is easy to show that
 \begin{align}
\label{eq:38}
\tilde \Pi_{11,\vec q}^{\text{el},R/A}(\omega) &= \tilde \Pi_{22,\vec q}^{\text{el},R/A}(\omega)   \\
\Pi_{12,\vec q}^{\text{el},R/A}(\omega) &= \Pi_{21,\vec q}^{\text{el},A/R}(-\omega) 
\label{cse11}  
 \end{align}
The Keldysh self-energies can be easily shown to be given by the thermal equilibrium expressions~\zref{nse8}.

 \subsection{Normal conductor}
In the normal conductor the propagators simplify to $\abs{P}_{v/c}=0$ and 
\begin{align}
\begin{split}
G_{0,\vec k,v/c}^{(p),R/A}(\omega) &= \frac{1}{\omega - \xi_{v/c}(\vec k) \pm i0}   \\
G_{0,\vec k,v/c}^{(p),K}(\omega) &= - 2 \pi i \bigl[1 - 2 n_F(\omega)\bigr] \delta\bigl[\omega-\xi_{v/c}(\vec k) \bigr]
\end{split}   \label{senc1}
\end{align}
where $n_F(\omega)$ is the Fermi function of the system, such that we find the standard results for the occurring particle-hole bubbles (the anomalous self-energies $\Pi_{11,\vec q}=\Pi_{22,\vec q}$ vanish in the normal conductor)
\begin{align}
\Pi_{21,\vec q}^{\text{el},R/A}(\omega) &= 2\abs{g_0^2} \sum_{\vec k}  \frac{n_F\bigl[\xi_v(\vec k)\bigr]-n_F\bigl[\xi_c(\vec k+\vec q)\bigr] }{\omega_- + \xi_v(\vec k) -\xi_c(\vec k+\vec q)\pm i0}  \label{senc2}
\end{align}
As depicted in Fig.~\ref{bandsnc}, let us now consider an isotropic electron band $\xi_c(\vec k) = \xi_c(\abs{\vec k})$ and a symmetric hole band $\xi_v(\vec k)=-\xi_c(\vec k)$ and linearize around the Fermi surface
\begin{align}
\xi_c(\vec k) \approx \vec v_F(\phi,\theta) \cdot \bigl[\vec k-\vec k_F(\phi,\theta)\bigr] = v_F (k-k_F)  \label{senc3}
\end{align}
where $\phi,\theta$ parametrize the orbital direction of $\vec k$ in spherical coordinates and we can define the DOS as usual $\rho(\xi) = \sum_{\vec k}  \delta\bigl[\xi-\xi_c(\vec k) \bigr]$. Let us now consider a finite momentum transfer with $q=\abs{\vec q} \ll k=\abs{\vec k} \approx \abs{\vec k_F}$. Due to the rotational invariance of the Fermi surface we can choose $\vec q$ to lie along the $z$ axis, see Fig.~\ref{Fermisurface}, such that
\begin{align}
\xi_c(\vec k+\vec q)&\approx \vec v_F(\phi,\theta) \cdot \bigl[\vec k+\vec q-\vec k_F(\phi,\theta)\bigr] = v_F (k-k_F) + \vec v_F(\phi,\theta) \cdot  \vec q = v_F (k-k_F) + v_F q \cos(\theta)   \nonumber \\
&= \xi_c(\vec k) + v_F q \cos(\theta)  \label{senc4}
\end{align}
which allows us to write the above sum of momenta as
\begin{figure}
\centering
\includegraphics[width=0.4\linewidth]{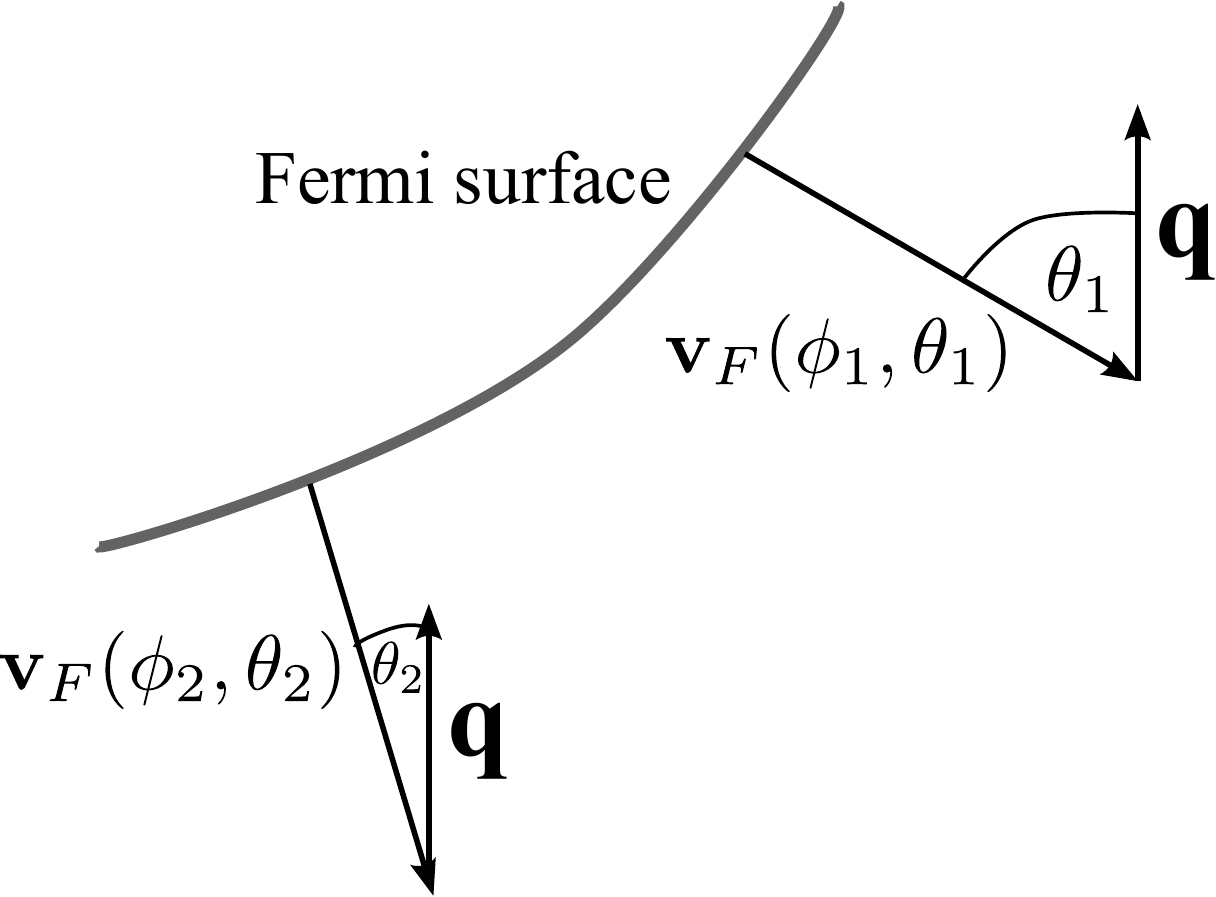}
\caption{Linearization of dispersion at the Fermi surface and finite momentum transfer by photon.}
 \label{Fermisurface}
\end{figure}
\begin{align}
\sum_{\vec k} f\bigl[\xi_v(\vec k),\xi_c(\vec k+\vec q)\bigr] &= \frac{1}{(2\pi)^3} \int_0^{2\pi} d\phi \int_{-1}^1 d\cos \theta \int_0^\infty dk \cdot k^2 \, f\bigl[-\xi_c(k),\xi_c(k)+v_F q \cos (\theta) \bigr]  \nonumber \\
&= \int_{-1}^1 \frac{d\cos \theta}2 \int_{-B}^\infty d\xi \, \rho(\xi)  f\bigl[ - \xi , \xi+ v_F q \cos(\theta) \bigr]  \nonumber \\
&=  \int_{-v_F q}^{v_F q} \frac{d\alpha}{2 v_F q}  \int_{-B}^\infty d\xi \, \rho(\xi)  f\bigl[ - \xi , \xi+ \alpha \bigr]   \label{senc5}
\end{align}
where $B$ is the energy distance from the lower band edge to the Fermi energy in the conduction band. We will now explicitly calculate the imaginary part of \zref{senc2} for the symmetric particle and hole bands in the normal conducting state and omit the real part since it is featureless and gives just a small unimportant correction to the dispersion of the photons. We find for zero fermionic temperature $T_F=0$ the imaginary part
\begin{align}
 \Im \Pi_{21,\vec q}^{\text{el},R/A}(\omega)  &=  \mp 2 \pi \abs{g_0^2}  \int_{-v_F q}^{v_F q} \frac{d\alpha}{2 v_F q}  \int_{-B}^\infty d\xi \, \rho(\xi) \biggl( n_F\bigl[-\xi\bigr]-n_F\bigl[\xi+\alpha\bigr]\biggr) \delta \bigl[\omega_- - 2\xi-\alpha  \bigr] \nonumber  \\
&= \mp \frac{ \pi \abs{g_0^2} }{2 v_F q} \int_{-v_F q}^{v_F q} d\alpha   \underbrace{\rho\bigl(\frac{\omega_- -\alpha}{2} \bigr)}_{\approx \rho(\omega_-/2)} \biggl( n_F\bigl[\frac{-\omega_- +\alpha}{2} \bigr]-n_F\bigl[\frac{\omega_- +\alpha}{2} \bigr]\biggr)   \nonumber \\
&\approx  \mp  \frac{ \pi \abs{g_0^2} }{2 v_F q}\rho(\omega_-/2) \int_{-v_F q}^{v_F q} d\alpha \biggl[ \theta(\omega_- -\alpha) - \theta(-\omega_- -\alpha)   \biggr]  \nonumber \\
&=  \mp  \pi \abs{g_0^2}\rho(\omega_-/2)  \, \sign(\omega_-) \, \min\bigl[1, \frac{\abs{\omega_-}}{v_F q} \bigr]      \label{senc6}
\end{align}
The real part will be neglected in the normal state as it is only weakly energy and momentum dependent and gives no important renormalization of the photon excitation energies $\omega_{\vec q}^* \approx \omega_{\vec q}$.

\subsection{Superconductor}  \label{appsup}
In the superconducting state we can use the standard parametrization of the normal and anomalous Greens functions ($T_F=0$)
\begin{align}
A_{0,\vec k,v/c}^{R/A}(\omega) &= \frac{\alpha_{A,\vec k,v/c}}{\omega-E_{v/c}(\vec k)\pm i0}+\frac{\beta_{A,\vec k,v/c}}{\omega-E_{v/c}(\vec k)\pm i0} \label{sesc1}  \\
A_{0,\vec k,v/c}^{K}(\omega)  &= \bigl[ 1- 2 n_F(\omega) \bigr] \biggl( A_{0,\vec k,v/c}^{R}(\omega)-A_{0,\vec k,v/c}^{A}(\omega)  \biggr) =- 2 \pi i \biggl( \alpha_{A,\vec k,v/c} \delta\bigl[\omega-E_{v/c}(\vec k) \bigr] +  \beta_{A,\vec k,v/c} \delta\bigl[\omega+E_{v/c}(\vec k) \bigr]    \biggr)   \nonumber
\end{align}
where we defined the superconducting dispersion $E_{v/c}(\vec k) = \sqrt{\xi_{v/c}(\vec k)^2 + \abs{\Delta_{v/c}}^2}$ and the coherence factors for the different Green's functions are given by
\begin{align}
\begin{array}{|c||c|c|c|c|}
\hline A & G^{(p)} & G^{(h)} & \abs{P}  \\ \hline \hline
\alpha_{A,\vec k,v/c} & u_{\vec k,c/v}^2 & v_{\vec k,c/v}^2  & -u_{\vec k,c/v} v_{\vec k,c/v}  \\ \hline
\beta_{A,\vec k,v/c} & v_{\vec k,c/v}^2  & u_{\vec k,c/v}^2  & u_{\vec k,c/v} v_{\vec k,c/v}   \\ \hline
\end{array}  \hspace{2cm}  \begin{matrix}
u_{\vec k,c/v} = \sqrt{ \frac{1}{2} \biggl( 1 + \frac{\xi_{c/v}(\vec k)}{E_{c/v}(\vec k)}\biggr)  }  \\
v_{\vec k,c/v} = \sqrt{ \frac{1}{2} \biggl( 1 - \frac{\xi_{c/v}(\vec k)}{E_{c/v}(\vec k)}\biggr)  }
\end{matrix}
 \label{sesc2}
\end{align}
An explicit calculation of the normal and anomalous self-energy show that
\begin{align}
\begin{split}
\tilde \Pi_{11,\vec q}^{\text{el},R/A}(\omega)&=   2\abs{g_0^2} \sum_{\vec k}  \biggl[ \frac{u_{\vec k,v} v_{\vec k,v}u_{\vec k+\vec q,c} v_{\vec k+\vec q,c}}{\omega-E_{v}(\vec k)-E_{c}(\vec k+\vec q)\pm i0} -\frac{u_{\vec k,v} v_{\vec k,v}u_{\vec k+\vec q,c} v_{\vec k+\vec q,c}}{\omega+E_{v}(\vec k)+E_{c}(\vec k+\vec q)\pm i0}           \biggr]  \\
\Pi_{21,\vec q}^{\text{el},R/A}(\omega)&=   2 \abs{g_0^2}\sum_{\vec k}  \biggl[ \frac{v_{\vec k,v}^2 u_{\vec k+\vec q,c}^2 }{\omega_--E_{v}(\vec k)-E_{c}(\vec k+\vec q)\pm i0} -\frac{u_{\vec k,v}^2 v_{\vec k+\vec q,c}^2}{\omega_-+E_{v}(\vec k)+E_{c}(\vec k+\vec q)\pm i0}           \biggr]    
\end{split}\label{sesc3}
\end{align}
and the other two self-energies are related via the symmetries~\zref{cse11}.

\subsubsection{Anomalous self-energy}
Let us now consider first the anomalous self-energy $\tilde \Pi_{11,\vec q}^{\text{el},R/A}(\omega)$ that can only occur in the presence of  superconducting quasi-particles. We can use the linearization \zref{senc5} for the symmetric bands to write the retarded self-energy as
\begin{align}
\tilde \Pi_{11,\vec q}^{\text{el},R}(\omega)&=   2 \abs{g_0^2} \int_{-v_F q}^{v_F q} \frac{d\alpha}{2 v_F q}  \int_{-B}^\infty d\xi \, \overbrace{\rho(\xi)}^{\approx \rho(0)=\rho_F} \frac{\abs{\Delta}^2}{4 \sqrt{\xi^2+\Delta^2}\sqrt{(\xi+\alpha)^2+\Delta^2}} \, \times \nonumber \\
&\hspace{5mm} \biggl[ \frac{1}{\omega-\sqrt{\xi^2+\Delta^2}-\sqrt{(\xi+\alpha)^2+\Delta^2}+ i0}  -\frac{1}{\omega+\sqrt{\xi^2+\Delta^2}+\sqrt{(\xi+\alpha)^2+\Delta^2}+ i0}           \biggr]   \nonumber \\
&= \frac{\abs{g_0^2} \rho_F \Delta}{4 v_F q} \int \limits_{-v_F q/\Delta}^{v_F q/\Delta} d\alpha  \int \limits_{-B/\Delta}^\infty dx \frac{1}{\sqrt{x^2+1}\sqrt{(x+\alpha)^2+1}}  \biggl[ \frac{1}{\tilde \omega-\sqrt{x^2+1}-\sqrt{(x+\alpha)^2+1} + i0} - \ldots     \biggr]   \label{sesc5a}
\end{align}
 where we assumed $\abs{\Delta_c}=\abs{\Delta_v}=\Delta$, defined $\tilde \omega=\omega/\Delta$ and approximated the DOS to lie near the Fermi surface since the dominant part of the integral will come from there. The real part of the integral is obviously symmetric and the imaginary part anti-symmetric in frequency
\begin{align}
\begin{split}
\Re \tilde \Pi_{11,\vec q}^{\text{el},R}(\omega)&=  \Re \tilde \Pi_{11,\vec q}^{\text{el},R}(-\omega)   \\
\Im \tilde \Pi_{11,\vec q}^{\text{el},R}(\omega)&=  - \Im \tilde \Pi_{11,\vec q}^{\text{el},R}(-\omega)
\end{split}  \label{sesc5b}
 \end{align} 
such that we will continue to focus on positive frequencies $\tilde \omega>0$. Let us first calculate the real part
\begin{align}
\Re \tilde \Pi_{11,\vec q}^{\text{el},R}(\omega)&= \frac{\abs{g_0^2} \rho_F \Delta}{4 v_F q} \mathcal P \int \limits_{-v_F q/\Delta}^{v_F q/\Delta} d\alpha  \int \limits_{-B/\Delta}^\infty dx \frac{1}{\sqrt{x^2+1}\sqrt{(x+\alpha)^2+1}}  \biggl[ \frac{1}{\tilde \omega-\sqrt{x^2+1}-\sqrt{(x+\alpha)^2+1} } +\bigl\{\omega \rightarrow - \omega \bigr\}    \biggr]   \label{sesc5c}
\end{align}
 for $\omega>0$. We see that the integrand will have a singularity if 
 \begin{align}
 \tilde \omega =\sqrt{x^2+1}+\sqrt{(x+\alpha)^2+1} \geq 2 \sqrt{1+ (\alpha/2)^2} =\tilde \omega_\alpha \hspace{1cm} , \text{for } x_{\text{min}}=-\alpha/2     \label{sesc5d}
 \end{align}
and we therefore expect that at $\tilde \omega=2$ some important feature can occur. Let us now consider this case $\omega \approx 2 \Delta$, where it is convenient to separate the singular region $x_{\text{min}} \approx - \alpha/2$ of the integrand
\begin{align}
\Re \tilde \Pi_{11,\vec q}^{\text{el},R}(\omega \approx 2 \Delta)&= g_{\vec q}(\omega) + \frac{\abs{g_0^2} \rho_F \Delta}{4 v_F q} \mathcal P \int \limits_{-v_F q/\Delta}^{v_F q/\Delta} d\alpha  \int \limits_{-\alpha/2-\delta x}^{-\alpha/2+\delta x} dx \frac{1}{\sqrt{x^2+1}\sqrt{(x+\alpha)^2+1}}  \frac{1}{\tilde \omega-\sqrt{x^2+1}-\sqrt{(x+\alpha)^2+1} } \nonumber \\
& \hspace{-4mm}\stackrel{x \approx - \alpha/2} \approx g_{\vec q}(\omega) + \frac{\abs{g_0^2} \rho_F \Delta}{4 v_F q} \mathcal P \int \limits_{-v_F q/\Delta}^{v_F q/\Delta} d\alpha  \int \limits_{-\alpha/2-\delta x \rightarrow - \infty}^{-\alpha/2+\delta x \rightarrow \infty} dx \frac{4}{\omega_\alpha^2}  \frac{1}{\tilde \omega-\tilde \omega_\alpha -\frac{8(x+\alpha/2)^2}{\tilde \omega_\alpha^{3}} }  \nonumber \\
&=  g_{\vec q}(\omega)- \frac{\pi \abs{g_0^2} \rho_F \Delta}{\sqrt{2}v_F q} \int \limits_{0}^{v_F q/\Delta} d\alpha  \frac{\theta(\tilde \omega_\alpha-\tilde \omega)}{\sqrt{\tilde \omega_\alpha} \sqrt{\tilde \omega_\alpha-\tilde \omega}}  \nonumber \\
&\approx g_{\vec q}(\omega) - \frac{\pi \abs{g_0^2} \rho_F \Delta}{v_F q} \ln \biggl[\sqrt{1+ \frac{ \bigl(\frac{v_F q}{2\Delta}\bigr)^2}{\abs{2-\tilde \omega}}}+ \frac{ \frac{v_F q}{2\Delta}}{\sqrt{\abs{2-\tilde \omega}}} \biggr]    \label{sesc5e}
\end{align}
The contribution $g_{\vec q}(\omega)$ will just give a small unimportant term as $\abs{g_0^2} \rho_F \ll \Delta$ is small and therefore our dispersion is just changed slightly. The important contributions will come from the divergences at $\omega = \pm 2 \Delta$, which are captured well by the expression \zref{sesc5e}, such that we approximate
\begin{align}
\Re \tilde \Pi_{11,\vec q}^{\text{el},R}(\omega)&\approx - \frac{\pi \abs{g_0^2} \rho_F }{2 \bigl( \frac{v_F q}{2\Delta}\bigr)} \ln \biggl[\sqrt{1+ \frac{ \bigl(\frac{v_F q}{2\Delta}\bigr)^2}{\abs{2-\abs{\tilde \omega}}}}+ \frac{ \frac{v_F q}{2\Delta}}{\sqrt{\abs{2-\abs{\tilde \omega}}}} \biggr]   \label{sesc5f}
\end{align} 

The imaginary part can  be calculated using the same methods as just explained for the real part and is given by
\begin{align}
\Im \tilde \Pi_{11,\vec q}^{\text{el},R}(\omega>0) &= - \frac{\abs{g_0^2} \pi \rho_F \Delta}{4 v_F q} \int \limits_{-v_F q/\Delta}^{v_F q/\Delta} d\alpha  \int dx \frac{1}{\sqrt{x^2+1}\sqrt{(x+\alpha)^2+1}} \delta \bigl[ \tilde \omega-\sqrt{x^2+1}-\sqrt{(x+\alpha)^2+1} \bigr]  \label{sesc5g}
\end{align}
Obviously, there is a gap of $2 \Delta$ for the imaginary part of the particle-hole bubble $\Im \tilde \Pi_{11,\vec q}^{\text{el},R}(\omega) \sim \theta(\omega-2 \Delta)$. Let us again focus on the region $\omega \gtrsim 2 \Delta$, where the dominant contributions come again from $x \approx -\alpha/2$ such that
\begin{align}
\Im \tilde \Pi_{11,\vec q}^{\text{el},R}(\omega \gtrsim 2 \Delta)& \approx  - \frac{\abs{g_0^2} \pi \rho_F \Delta}{4 v_F q} \int \limits_{-v_F q/\Delta}^{v_F q/\Delta} d\alpha  \int dx \frac{4}{\tilde \omega_\alpha^2}\delta \biggl[ \tilde \omega-\tilde \omega_\alpha -\frac{8(x+\alpha/2)^2}{\tilde \omega_\alpha^{3}} \biggr]    \nonumber\\
&= - \frac{\abs{g_0^2} \pi \rho_F \Delta}{2\sqrt{2} v_F q} \int \limits_{-v_F q/\Delta}^{v_F q/\Delta} d\alpha  \int dy \frac{1}{\sqrt{\tilde \omega_\alpha}} \delta \bigl[ \tilde \omega-\tilde \omega_\alpha -y^2 \bigr] \nonumber \\
&= - \frac{\abs{g_0^2} \pi \rho_F \Delta}{2\sqrt{2} v_F q} \int \limits_{-v_F q/\Delta}^{v_F q/\Delta} d\alpha  \frac{\theta(\tilde \omega-\tilde \omega_\alpha)}{\sqrt{\tilde \omega_\alpha}\sqrt{\tilde \omega-\tilde \omega_\alpha}} \approx - \frac{\abs{g_0^2} \pi \rho_F}{2 \bigl( \frac{v_F q}{2\Delta}\bigr)} \arcsin\biggl[ \frac{\min \bigl[\frac{v_F q}{2\Delta},\sqrt{\tilde \omega-2} \bigr]}{\sqrt{\tilde \omega - 2}}   \biggr]   \label{sesc5h} 
\end{align}
where in the end we approximated the integral for small $\alpha$ due to the integrand restrictions. Far away from the gap region $\omega \gg 2 \Delta$ it is easy to show that the imaginary part $\Im \tilde \Pi_{11,\vec q}^{\text{el},R}(\omega >> 2 \Delta) \sim 1/\omega^2$ falls down rapidly and is therefore not important, such that we can approximate $\Im \tilde \Pi_{11,\vec q}^{\text{el},R}(\omega)$ for all $\omega$ with \zref{sesc5h} as stated in the main text.

\subsubsection{Normal self-energy}
The normal self-energy $\Pi_{21,\vec q}^{\text{el},R/A}(\omega)$ around $\omega_- =\omega- e V_0= \pm 2 \Delta$ can easily be shown to behave exactly like the anomalous self-energy $\tilde \Pi_{11,\vec q}^{\text{el},R/A}(\omega)$ around $\omega = \pm 2 \Delta$. Far away from this region $\abs{\omega_-} \gg 2 \Delta$ it just behaves like in the normal conductor \zref{senc6}. In between, we interpolate between these two by replacing the self-energy as the maximum of the two limits
\begin{align}
\Pi_{21,\vec q}^{\text{el},R/A}(\omega)& \approx - \frac{\pi \abs{g_0^2} \rho_F }{2 \bigl( \frac{v_F q}{2\Delta}\bigr)} \ln \biggl[\sqrt{1+ \frac{ \bigl(\frac{v_F q}{2\Delta}\bigr)^2}{\abs{2-\abs{\tilde \omega}}}}+ \frac{ \frac{v_F q}{2\Delta}}{\sqrt{\abs{2-\abs{\tilde \omega}}}} \biggr]   \nonumber \\
& \quad  \mp i \max\biggl[ \frac{\abs{g_0^2} \pi \rho_F}{2 \bigl( \frac{v_F q}{2\Delta}\bigr)} \arcsin\biggl( \frac{\min \bigl[\frac{v_F q}{2\Delta},\sqrt{\abs{\tilde \omega}-2}\bigr] }{\sqrt{\abs{\tilde \omega}- 2}}   \biggr) , \pi \abs{g_0}^2 \rho(\omega_-/2)    \biggr] \sign(\omega_-) \theta(\abs{\tilde \omega}-2)
\end{align}

\section{Derivation of the dressed propagators}\label{appendixdressedpropagators}
Let us first consider the Wigner transformation of a convolution of two bosonic matrices $ \ring C = \ring A \circ \ring B$ with time dependence
\begin{align}
\begin{matrix} \ring A(t,t') = \mx{ e^{i\phi(2T)} A_{11}(\tau)   &  A_{12}(\tau)  \\  A_{21}(\tau)  & e^{-i\phi(2T)} A_{22}(\tau) } \\ \ring B(t,t') = \mx{ e^{-i\phi(2T)} B_{11}(\tau)   &  B_{12}(\tau)  \\  B_{21}(\tau)  & e^{i\phi(2T)} B_{22}(\tau) } \end{matrix}  \hspace{2cm}   \begin{matrix} \ring A(\omega,T) = \mx{ e^{i\phi(2T)} A_{11}(\omega)   &  A_{12}(\omega)  \\  A_{21}(\omega)  & e^{-i\phi(2T)} A_{22}(\omega) } \\ \ring B(\omega,T) = \mx{ e^{-i\phi(2T)} B_{11}(\omega)   &  B_{12}(\omega)  \\  B_{21}(\omega)  & e^{i\phi(2T)} B_{22}(\omega) } \end{matrix}     \label{dp1}
\end{align}
where as usual we set $\phi(2T)=2 e V_0 T - \phi_c+\phi_v + 2 \phi_g$, defined $T=\frac{t+t'}{2}$ and $\tau=t-t'$. The Wigner-transform of a convolution $ \ring C = \ring A \circ \ring B$ can be expressed conveniently by the Wigner transforms of $\ring A$ and $\ring B$ as
\begin{align}
\ring C(\omega,T) = \ring A(\omega,T) e^{ \frac{i}{2} \bigl[ \overleftarrow \partial_T \overrightarrow \partial_\omega - \overleftarrow \partial_\omega \overrightarrow \partial_T   \bigr]  } \ring B(\omega,T)   \label{dp2}
\end{align}
Using the identities
\begin{align}
\begin{split}
e^{\frac{i}{2} a \cdot \partial_T} e^{\pm i \phi(2T)} &= e^{\mp a \cdot  e V_0}  e^{i \phi(2T)} \\
e^{ \pm a \partial_\omega} f(\omega) &= f(x \pm a)   
\end{split}  \label{dp3}
\end{align}
we get
\begin{align}
\ring C(\omega,T)&= \mx{ e^{i\phi(2T)} A_{11}(\omega)   &  A_{12}(\omega)  \\  A_{21}(\omega)  & e^{-i\phi(2T)} A_{22}(\omega) } e^{ \frac{i}{2} \bigl[ \overleftarrow \partial_\omega \overrightarrow \partial_T - \overleftarrow \partial_T \overrightarrow \partial_\omega   \bigr]  } \mx{ e^{-i\phi(2T)} B_{11}(\omega)  &  B_{12}(\omega)  \\  B_{21}(\omega)  & e^{i\phi(2T)} B_{22}(\omega)  }    \nonumber \\
&= \mx{ e^{i\phi(2T)} A_{11}(\omega) e^{ -\frac{i}{2}  \overleftarrow \partial_T \overrightarrow \partial_\omega}   &  A_{12}(\omega)  \\  A_{21}(\omega)  & e^{-i\phi(2T)} A_{22}(\omega) e^{- \frac{i}{2}  \overleftarrow \partial_T \overrightarrow \partial_\omega}} \mx{ e^{ \frac{i}{2} \overleftarrow \partial_\omega \overrightarrow \partial_T   }  e^{-i\phi(2T)} B_{11}(\omega)   &  B_{12}(\omega)  \\  B_{21}(\omega)  & e^{ \frac{i}{2} \overleftarrow \partial_\omega \overrightarrow \partial_T   }  e^{i\phi(2T)} B_{22}(\omega) }    \nonumber \\
&= \mx{ e^{i\phi(2T)} A_{11}(\omega) e^{ e V_0 \overrightarrow \partial_\omega}   &  A_{12}(\omega)  \\  A_{21}(\omega)  & e^{-i\phi(2T)} A_{22}(\omega) e^{ -e V_0 \overrightarrow \partial_\omega}} \mx{ e^{ e V_0 \overleftarrow \partial_\omega}   e^{-i\phi(2T)} B_{11}(\omega)   &  B_{12}(\omega)  \\  B_{21}(\omega)  & e^{ -e V_0 \overleftarrow \partial_\omega}  e^{i\phi(2T)} B_{22}(\omega) }  \nonumber \\
&= \mx{ A_{11}(\omega+ e V_0) B_{11}(\omega+eV_0)   + A_{12}(\omega)   B_{21}(\omega) &  e^{i \phi(2T)} \bigl[ A_{11}(\omega)   B_{12}(\omega+ e V_0)+ A_{12}(\omega- e V_0)   B_{22}(\omega) \bigr] \\  e^{-i \phi(2T)} \bigl[ A_{21}(\omega+ e V_0)   B_{11}(\omega)+ A_{22}(\omega)   B_{21}(\omega- e V_0) \bigr] &  A_{21}(\omega) B_{12}(\omega)   + A_{22}(\omega- e V_0)   B_{22}(\omega- e V_0)  }   \label{dp4}
\end{align}
The Dyson equation in Keldysh space is formally given by the convolution
\begin{align}
\bigl( \hat D_{0,\vec q}^{-1}-\hat \Pi_{\vec q} \bigr) \circ \hat D_{\vec q}^K &= \hat {\openone} \, ,   \label{dp5}
\end{align}
where $\hat \Pi_{\vec q}= \hat \Pi_{\vec q}^{\text{el}}+\Pi_{\vec q}^{\text{bath}}$ is the full self-energy of the system. The three coupled equations for the full retarded, advanced and Keldysh propagators of the photon
\begin{align}
\begin{split}
\bigl( [\ring D_{0,\vec q}^{R/A}]^{-1}-\ring \Pi_{\vec q}^{R/A} \bigr)\circ \ring D_{\vec q}^{R/A} &= \ring 1 \, , \\
\bigl( [\ring D_{0,\vec q}^{R}]^{-1}-\ring \Pi_{\vec q}^{R} \bigr) \circ \ring D_{\vec q}^K &= \ring \Pi_{\vec q}^{K} \circ \hat D_{\vec q}^A \, .
\end{split}   \label{dp6}
\end{align}
can now be solved with the ansatz that
\begin{align}
\ring D_{\vec q}^{R,A,K}(\omega,T) &= \mx{e^{- i\phi(2T)}\tilde D_{11,\vec q}^{R,A,K}(\omega) & D_{12,\vec q}^{R,A,K}(\omega)   \\ d_{21,\vec q}^{R,A,K}(\omega) & e^{ i\phi(2T)} \tilde D_{22,\vec q}^{R,A,K}(\omega)}  \, ,\label{dp7}
\end{align}
has the structure of $\ring B$. Since  both
\begin{align}
\begin{split}
\bigl( [\ring D_{0,\vec q}^{R}]^{-1}(\omega,T)-\ring \Pi_{\vec q}^{R} (\omega,T)\bigr) &= \mx{ - e^{i \phi(2T)} \tilde \Pi_{11,\vec q}^{R}(\omega) &  \overbrace{-\omega-\omega_{\vec q} - i0}^{[d_{0,\vec q}^A(-\omega)]^{-1}}-  \Pi_{12,\vec q}^{R}(\omega) \\ \underbrace{\omega-\omega_{\vec q} +i0}_{[d_{0,\vec q}^R(\omega)]^{-1}}- \Pi_{21,\vec q}^{R}(\omega) & -e^{-i \phi(2T)} \tilde \Pi_{22,\vec q}^{R}(\omega) }  \\
\ring \Pi_{\vec q}^{K} (\omega,T) &= \mx{  e^{i \phi(2T)} \tilde \Pi_{11,\vec q}^{K}(\omega) &   \Pi_{12,\vec q}^{K}(\omega) \\  \Pi_{21,\vec q}^{K}(\omega) & e^{-i \phi(2T)} \tilde \Pi_{22,\vec q}^{K}(\omega) }
\end{split}   \label{dp8} 
\end{align}
have exactly the structure of $\ring A$ we can use \zref{dp4} to calculate the Wigner transform of the convolutions in \zref{dp6} exactly, which leads to the propagators defined in \zref{prop}.

\end{widetext}


%

\end{document}